\documentclass[pdflatex,sn-mathphys-num]{sn-jnl}

\usepackage[utf8]{inputenc}
\usepackage{graphicx}%
\usepackage{multirow}%
\usepackage{amsmath,amssymb,amsfonts}%
\usepackage{amsthm}%
\usepackage{mathrsfs}%
\usepackage[title]{appendix}%
\usepackage{xcolor}%
\usepackage{textcomp}%
\usepackage{manyfoot}%
\usepackage{booktabs}%
\usepackage{verbatim}
\usepackage{algorithm}%
\usepackage{algorithmicx}%
\usepackage{algpseudocode}%
\usepackage{listings}%
\usepackage[normalem]{ulem}
\usepackage[switch]{lineno}

\theoremstyle{thmstyleone}%
\theoremstyle{thmstyletwo}%

\theoremstyle{thmstylethree}%

\begin{document}

\title[Real-time reinforcement learning for turbulent state-dependent control in a bluff-body wake]{Real-time reinforcement learning for turbulent state-dependent control in a bluff-body wake}


\author[1]{\fnm{Junjie} \sur{Zhang}}\email{jacky.zhang20@imperial.ac.uk}

\author[1]{\fnm{Chengwei} \sur{Xia}}\email{chengwei.xia20@imperial.ac.uk}

\author[1]{\fnm{Xianyang} \sur{Jiang}}\email{x.jiang@imperial.ac.uk}

\author[1]{\fnm{Isabella} \sur{Fumarola}}\email{isabella.fumarola12@imperial.ac.uk}

\author*[1]{\fnm{Georgios} \sur{Rigas}}\email{g.rigas@imperial.ac.uk}

\affil[1]{\orgdiv{Department of Aeronautics}, \orgname{Imperial College London}, \orgaddress{\city{London}, \postcode{SW7 2AZ}, \country{UK}}}


\abstract{
\textcolor{black}{Controlling turbulent dynamics remains a major challenge because of its chaotic, multi-scale dynamics, which strongly influence the performance of many fluid systems.}  Here we report REACT (Reinforcement Learning for Environmental Adaptation and Control of Turbulence), \textcolor{black}{an autonomous reinforcement learning framework for real-time state-dependent control of turbulent wake dynamics in a real wind-tunnel environment.} 
Deployed on a \textcolor{black}{Ahmed-body model} equipped solely with onboard sensors and servo-actuated surfaces, REACT learns directly from sparse experimental measurements in a wind-tunnel environment, bypassing empirical turbulence models. The agent autonomously converges to a policy that reduces aerodynamic drag while achieving net energy savings. Without prior knowledge of flow physics, it discovers that dynamically suppressing spatiotemporally coherent flow structures in the \textcolor{black}{bluff-body} wake maximizes energy efficiency, achieving two to four times greater performance than model-based baseline controllers. We contrast REACT’s state-dependent, dynamics-aware policy with representative quasi-steady, mean-flow-oriented policies learned by standard reinforcement learning baselines, which deliver lower drag reduction and no direct suppression of coherent instabilities in this turbulent-wake regime. Finally, by training in a nondimensional state–reward space whose amplitudes are approximately Reynolds-number-invariant, and by conditioning on Reynolds number for temporal adaptation, REACT learns a single offline policy that remains effective across the tested Reynolds-number range \(\text{\normalfont Re = 86{,}400 to 518{,}400}\), without retraining.
\textcolor{black}{These results demonstrate autonomous closed-loop reinforcement learning control in a high-Reynolds-number wind-tunnel environment and suggest a path toward data-driven state-dependent control of turbulent flows.}}





\maketitle

\section{Introduction}\label{sec1}

The chaotic and multi-scale motion of fluids in space and time, known as turbulence, governs the transfer of energy between objects and the surrounding fluid. Despite decades of research focused on the prediction and characterization of turbulent behavior~\citep{feynman1964lectures} spanning the transition to turbulence~\citep{hof2006finite, barkley2015rise, shih2016ecological} and the emergence of turbulent patterns~\citep{reetz2019exact, huisman2014multiple,callaham2022empirical} to the cascade of energy across scales~\citep{de2024pattern, young2017forward}, effective methods for controlling turbulence remain elusive. In engineering applications, control of turbulent dynamics~\citep{brunton2015closed} can be an enabler of transformative advances in all systems that interact with a fluid, with implications ranging from energy-efficient transport~\citep{marusic2021energy} to enhancing renewable energy harvesting~\citep{shapiro2022turbulence}.

Flow control strategies in turbulent regimes, at increasing levels of control authority and design complexity, include passive, open-loop, and closed-loop approaches~\citep{choi2008control}. Passive methods rely on geometric shape optimization, whereas open-loop approaches rely on predefined actuation. Both modify turbulence production and energy transfer indirectly through mean-flow modifications rather than directly manipulating the dynamic behavior of turbulence, rendering their performance suboptimal to closed-loop approaches. Due to the complexity of turbulent flows, the model-based design of closed-loop controllers has been restricted to the linear regime~\citep{kim2007linear,jovanovic2021bypass}, limiting their ability to address the non-linear, multi-scale, high-Reynolds-number physics of flows of practical interest.

Reinforcement learning (RL) has recently enabled model-free closed-loop control across a range of complex physical systems, including drone racing~\citep{kaufmann2023champion}, robotics control~\citep{han2024lifelike, radosavovic2024real, andrychowicz2020learning, lee2020learning} and tokamak plasma control~\citep{degrave2022magnetic}. Turbulence, however, presents a fundamentally harder challenge: its dynamics are chaotic and formally infinite-dimensional,  governed by the Navier–Stokes equations~\citep{pope2000turbulent}, yet are practically approximated as extremely high-dimensional. Therefore, simulation-to-reality transfer~\citep{tobin2017domain}, widely adopted in robotics~\citep{kaufmann2023champion, han2024lifelike, radosavovic2024real, andrychowicz2020learning, lee2020learning} or plasma~\citep{degrave2022magnetic} control, remains highly challenging for high-Reynolds-number turbulent flows. Furthermore, the spatiotemporal turbulent dynamics evolve non-linearly and non-locally resulting in rich multi-scale and chaotic interactions that challenge AI agents to navigate through the complex phase-space during training and thus typically converge to suboptimal strategies independent of the turbulent state (i.e. static or open-loop strategies). Beyond the intrinsic complexity of turbulence, the limited spatiotemporal sensor resolution and practical placement leave the turbulent flow state only partially observed, creating a strongly non-Markovian control problem~\citep{kaelbling1998planning} that challenges the convergence of RL to optimal solutions.

Recent RL-based flow-control studies have demonstrated promising performance at tractable Reynolds numbers in simulation environments \cite{pivot2017continuous,tang2020robust,paris2021robust,guastoni2023deep,chen2023deep,font2025deep,wang2024learn,wang2024dynamic,xia2024active,sonoda2023reinforcement,ren2021applying,rabault2019artificial,verma2018efficient}. Experimental demonstrations are also promising, including gust rejection and separation control around wings \cite{renn2022machine,shimomura2020closed}, mixing enhancement in supersonic flows \cite{zong2025closed} and attenuation of vortex-induced vibrations~\citep{chen2023deep}.  For turbulent wake suppression aimed at drag reduction, model-free machine-learning optimizers have been successfully employed, such as genetic-programming–driven control and explorative-gradient–method optimization \cite{gautier2015closed,debien2016closed,li2017drag,fan2020optimization}, as well as RL-based approaches \cite{fan2020reinforcement,dong2023surrogate, amico2022deep,amico2024flow}. These seminal studies efficiently realize mean-flow–oriented feedback strategies: the controller adjusts around a discovered optimal periodic or quasi-static actuation so that coherent instabilities are mitigated indirectly through their dependence on the mean flow, yielding compact and widely applicable controller designs. By contrast, model-based controllers can realize genuinely dynamical, state-dependent strategies that act directly on the hydrodynamic instabilities \citep{brackston2016stochastic,pastoor2008feedback, li2016feedback, ahmed2022nonlinear, brackston2018modelling} However, these approaches face an inherent trade-off: suppressing one instability often amplifies others. For example, Li et al.\citep{li2016feedback} achieved wake symmetrization but enhanced vortex shedding, limiting pressure recovery to 2\%; Brackston et al.\citep{brackston2018modelling} suppressed bistability but amplified higher-frequency dynamics, precluding net drag reduction. These studies highlight the challenge of multi-modal turbulent control: without a principled way to balance competing instabilities, model-based designs remain suboptimal.  In this study, \textcolor{black}{we demonstrate experimental RL agent (model-free) to exhibit dynamical and state-dependent behavior while actively targeting multi-scale turbulent dynamics.}
At the same time, robust \textcolor{black}{extrapolation} across varying speed conditions and the extraction of interpretable physical strategies remain open challenges in both turbulence control and scientific AI more broadly \cite{brunton2020machine,kirk2023survey,li2023critical}.

In this work, we present Reinforcement Learning for Environmental Adaptation and Control of Turbulence (REACT), a real-time RL-based platform for closed-loop control of turbulent wake dynamics, and experimentally validate its performance in the fully turbulent wake of a Ahmed-body model~\cite{ahmed1984some} in a wind-tunnel experiment at turbulent Reynolds numbers ($Re$) from 86,400 to 518,400. Rather than addressing a flow separation/reattachment problem, the current configuration has a fixed separation topology, and REACT acts on the already separated turbulent wake to regulate the turbulent dynamics. REACT learns efficiently from sparse, on-body pressure measurements to dynamically suppress spatio-temporal coherent structures and autonomously achieve net energy savings. By \emph{autonomous}, we mean the feedback law was discovered model-free via learning directly from real-time experimental interaction--without human intervention, and crucially, without relying on prescribed reduced-order models, observers, or hand-tuned control structures. Furthermore, with a physics-consistent scaling and Re conditioning (i.e., expressing the state and reward as the nondimensional coefficients $C_d$ and ${C}_p$, which are  asymptotically Re-invariant in fully turbulent, pressure-drag–dominated wakes, and conditioning on Re to accommodate temporal-scale shifts) \textcolor{black}{REACT learns a single, offline-trained controller that remains effective across the range of wind speeds tested. Together, these elements make REACT an autonomous, dynamics-aware controller that delivers aerodynamic drag reduction and net energy savings across the operating conditions tested in wind-tunnel environment.}

\section{The REACT system}
\label{sec:react system}

\begin{figure}
    \centering
    \includegraphics[width=1\linewidth]{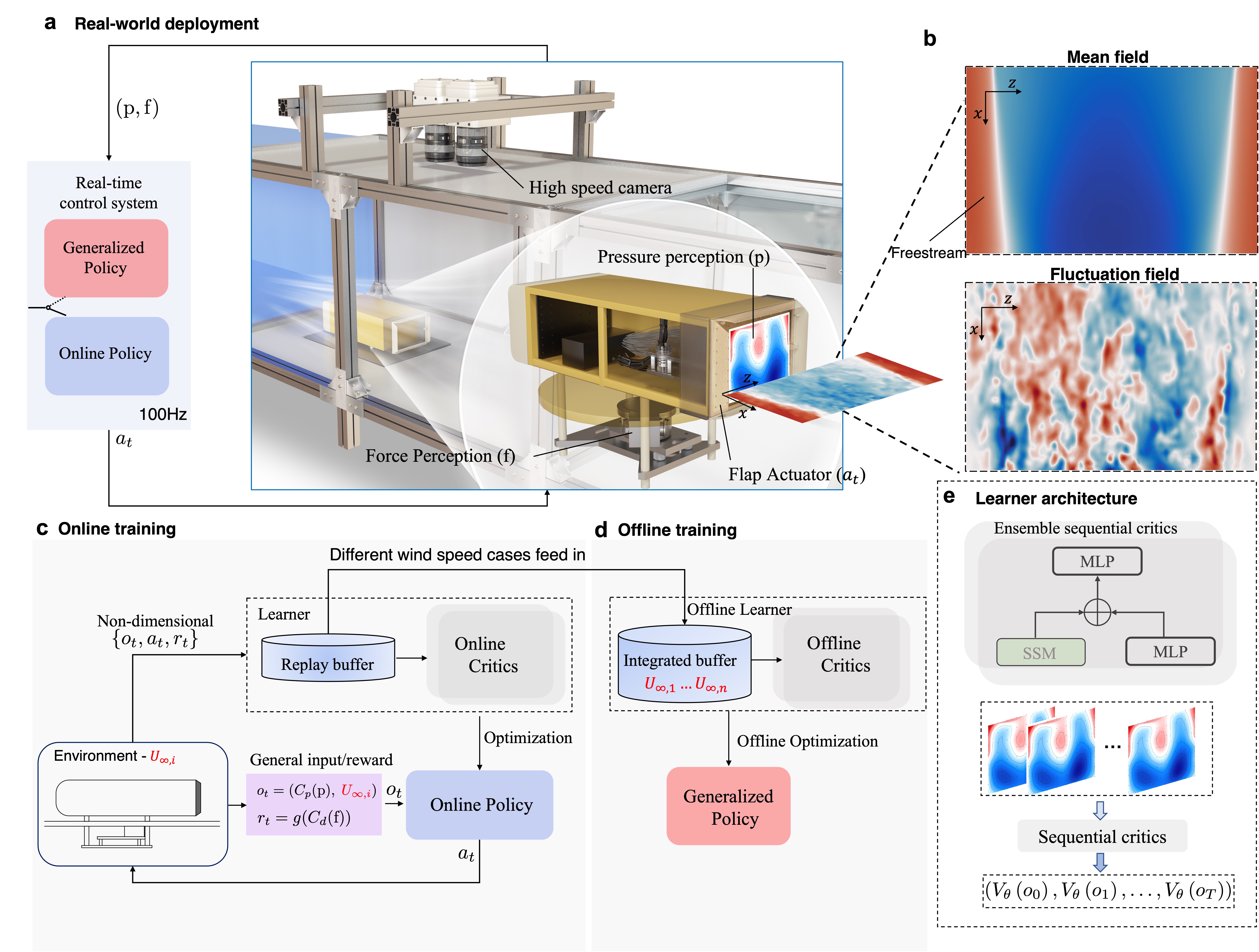}
    
    \caption{\textbf{The REACT system.} (\textbf{a}) Deployment of the REACT system with a real-time feedback loop between the agent and wind-tunnel environment. The relative speed of the vehicle to the incoming flow is $U_\infty$. The pressure field is sensed via on-vehicle base-pressure taps; a load cell is measuring the aerodynamic forces, and velocity flow fields are captured by a ceiling-mounted camera. The rear flap actuators enable active flow control. (\textbf{b}) Mean velocity and instantaneous fluctuation velocity field. (\textbf{c}) Online learning loop of the RL algorithm. (\textbf{d}) Offline training loop. (\textbf{e}) Network architecture of the critics within the learning loop.}
    \label{fig:REACT}
\end{figure}

The REACT system (Fig.~\ref{fig:REACT}) is a learning-based platform for real-time control of turbulent wake dynamics, developed to improve aerodynamic performance under variations in wind speed. During the wind-tunnel experiments, the model was subjected to a uniform freestream velocity $U_{\infty}$ in the reference frame of the laboratory. REACT integrates two key components: (i) a real-time control loop that generates time-critical actuation commands from the RL policy based on instantaneous feedback from onboard sensors; and (ii) a training loop that continually updates the policy to optimize energy savings. To assess turbulence suppression, the velocity field is captured by particle image velocimetry (PIV). In the uncontrolled state, the wake exhibits broadband unsteadiness and multi-scale flow structures (Fig.~\ref{fig:REACT}(b)) that contribute to aerodynamic resistance.

The real-time control loop (Fig.~\ref{fig:REACT}(a)) links the perception unit, control  policy, and actuators in a low-latency feedback architecture. The perception unit consists of a high-speed pressure scanner and a load cell that infer the turbulent wake state and measure aerodynamic forces, respectively. The scanner, embedded within the vehicle body, samples signals from 64 on-vehicle pressure taps distributed across its rear base. Real-time aerodynamic forces and actuator energy consumption form the basis of the reward signal guiding the RL agent towards energy-saving strategies. Flow control is achieved through two servo-actuated flaps mounted vertically at the rear edges, which are driven in real time by commands from the GPU-hosted RL policy, which communicates with the control system via UDP~\cite{postel1980user}.

The training framework (Fig.~\ref{fig:REACT}(c),(d)) integrates  three components to address the challenges of turbulent flow control under varying conditions: (i) online and offline training, designed to enable policy generalization; (ii) a critic module for value~\cite{sutton2018reinforcement} estimation based on an ensemble, sequence-to-sequence state-space model (SSM), which mitigates stochasticity, partial observability, and high dimensionality in turbulent prediction; and (iii) both online and offline policies adopt a branched multi-layer perceptron–long short-term memory (MLP–LSTM) architecture to handle the challenges of partial observability inherent in turbulence control. The online training loop is adapted from the Soft Actor-Critic (SAC) framework~\cite{haarnoja2018soft}.

Extrapolating RL agents to operate effectively under parametric variations (here changes in \(U_\infty\) and the corresponding \(Re\)) is challenging due to out-of-distribution states. One route to extrapolate is to exploit invariances of the underlying spatiotemporal dynamics and learn a policy that is invariant (or appropriately equivariant) to parametric changes. Here, we exploit invariances \emph{and} context: in high-\(Re\) turbulent wakes, the Taylor–Kolmogorov “zeroth law” (consistent with K41) implies that the nondimensional dissipation is approximately \(Re\)-independent~\citep{Kolmogorov1941c,pope2000turbulent}; together with a global kinetic-energy balance for a statistically steady wake, this yields order-one, weakly \(Re\)-dependent force/pressure coefficients (e.g., \(C_d\), \(C_p\))~\citep{roshko1961experiments}. We therefore train in a nondimensional state–reward space and explicitly condition on \(Re\) to accommodate temporal-scale shifts. Accordingly, REACT trains the \emph{online policy} on \((C_p, C_d)\) with an \(Re\) context (Methods~\ref{sec:Methods_generalization}), preserving approximately invariant amplitudes across speeds while learning temporal adaptation; trajectories are logged to a replay buffer for offline updates, yielding a single policy that transfers across speeds without retraining.

Turbulent flows are deterministic under the Navier–Stokes equations, yet to the perception unit in real-world environments appear stochastic due to unresolved fine-scale dynamics and environmental disturbances. This apparent stochasticity corrupts value estimation, inflating predicted returns in ways that slow convergence and can trap the policy in suboptimal local minima. REACT addresses this by employing an ensemble of independently initialized critic networks (Fig.~\ref{fig:REACT}(e)) and adopting a conservative, minimum-based aggregation of their predictions. This approach suppresses overestimation bias, enhances robustness to noisy or inaccurate returns, and achieves stable learning in turbulent environments.

\textcolor{black}{The partially observed nature of turbulent wake control substantially elevates the learning problem beyond standard fully observed RL. In our setting, the agent has access only to rear-surface pressure measurements, rather than the full flow field, reflecting realistic sensing constraints in experimental aerodynamics \cite{xia2024active}. The control problem is therefore naturally formulated as a partially observable Markov decision process (POMDP) \cite{kaelbling1998planning}, in which effective decision-making requires the critic to reconstruct a belief-like representation of the hidden wake state from histories of observations and actuation. This requirement is particularly acute in turbulent flows, where energetically important structures evolve over multiple coupled timescales, exhibit intermittent switching, and leave only indirect signatures in pressure measurements. To address this, the ensemble critic combines an instantaneous feedforward pathway with a sequence-model pathway, allowing value estimation to exploit both local observation features and latent dynamical context. For the latter, we adopt a selective state-space model based on the Mamba architecture \cite{gu2023mamba} within a sequence-to-sequence critic framework (Fig.~\ref{fig:REACT}(e); Methods~\ref{sec:critic design}). The selective SSM enables the critic to retain, suppress, or propagate temporal information according to the evolving flow condition. This makes it well suited to episodic sequential value estimation, where coherent structures and slow recovery processes are only weakly observable from instantaneous measurements, but become identifiable over longer temporal horizons. Empirically, we find that the Mamba-based critic ensemble converges with fewer gradient updates and supports the discovery of sustained energy-saving policies, whereas the LSTM-based alternative fails to do so reliably. Ablations over critic architecture and ensemble size (Methods~\ref{sec:critic design}) further show that both pessimistic ensemble aggregation and selective sequence modeling are critical to REACT’s performance in this chaotic, partially observable control regime.}

\section{Dynamic, state-dependent versus mean-flow control}

\begin{figure}
    \centering
    \includegraphics[width=\linewidth]{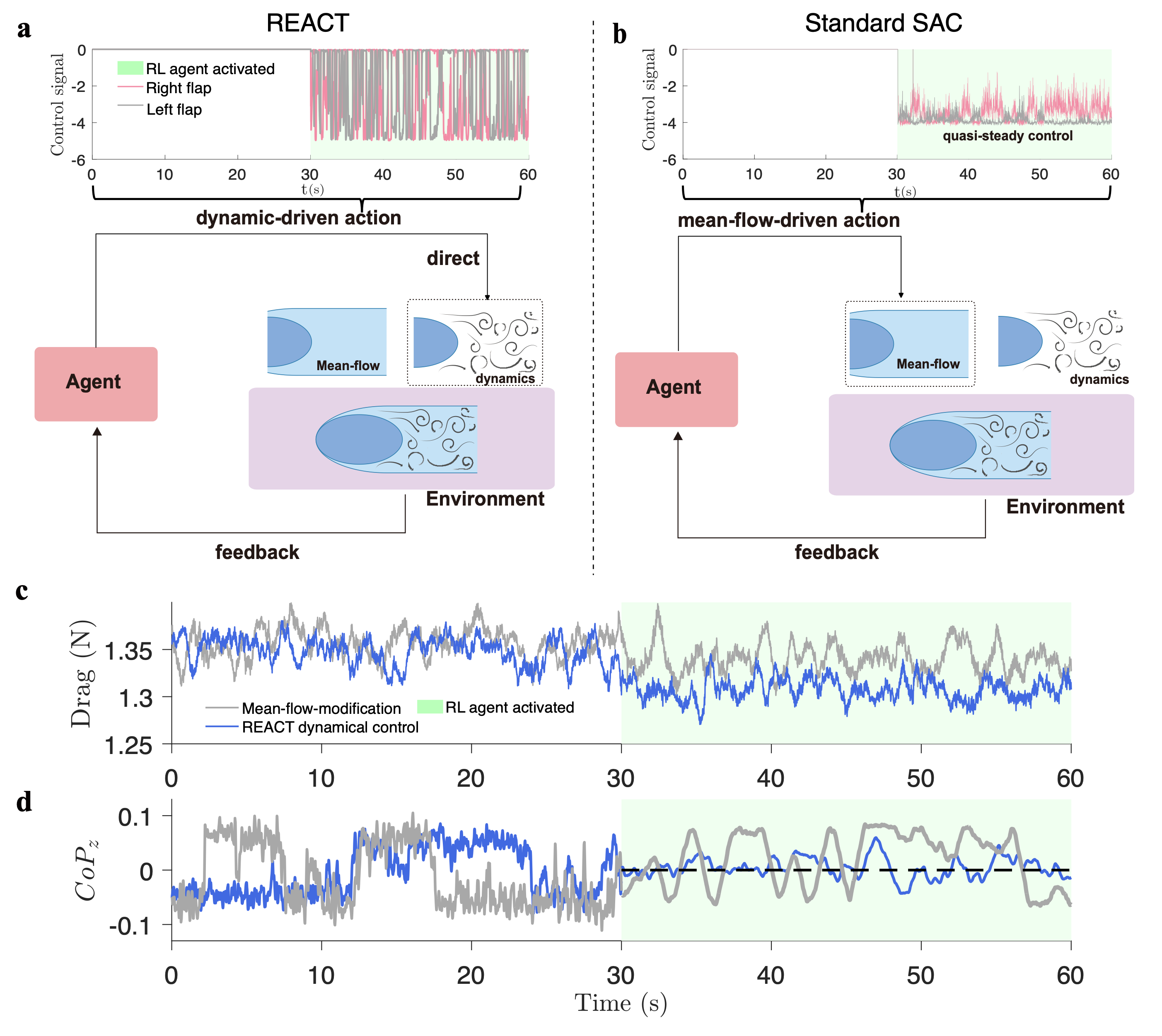}
    \caption{\textbf{Dynamics-aware versus mean-flow–oriented control.}
    (\textbf{a}) REACT (SSM critic ensemble) learns \emph{dynamics-aware} control: commands remain state-dependent and target unsteady instabilities.
    (\textbf{b}) A vanilla SAC baseline (MLPs for actor/critic networks) converges to a \emph{mean-flow–oriented} quasi-steady strategy with nearly symmetric steady deflections.
    (\textbf{c}) Aerodynamic drag traces (0.6\,s moving mean) showing stronger reduction with REACT.
    (\textbf{d}) Base center-of-pressure $CoP_{z}$ (1\,s moving mean during control), indicating suppression of instability under REACT.}
    \label{fig:dynamical_meanflow}
\end{figure}

In partially observed, stochastic turbulent flows, model-free RL can converge to two qualitatively different regimes (Fig.~\ref{fig:dynamical_meanflow}a,b), depending on whether the function approximator can capture the \emph{temporal} evolution of the turbulent state and assign credit to state-dependent actions.

\emph{Dynamics-aware control.} In the first regime (Fig.~\ref{fig:dynamical_meanflow}a), the agent targets unsteady coherent structures directly. Within REACT, a sequential \emph{state-space model} (SSM) critic \emph{ensemble} provides the temporal memory and robustness needed under partial observability (POMDP), mitigating value overestimation from noisy observations. The learned policy remains tightly coupled to the instantaneous pressure field: flap commands are strongly correlated/coherent with $C_p(t)$, yielding a \emph{dynamic}, state-dependent law that reduces drag (Fig.~\ref{fig:dynamical_meanflow}c) and suppresses coherent instabilities (Fig.~\ref{fig:dynamical_meanflow}d). Consistent with closed-loop regulation rather than quasi-steady scheduling, \emph{open-loop replay} (see Methods~\ref{sec:closeloop_strategy_validation}) of the same actuation sequence fails to reproduce the benefit. Mechanistically, it achieves strong suppression of coherent instabilities (Fig.~\ref{fig:dynamical_meanflow}d), as shown in the instability proxy, given by center of pressure $CoP_z$ at the base, which is also described in more detail in the next sections.

\emph{Mean-flow–oriented control.} In the second regime (Fig.~\ref{fig:dynamical_meanflow}b), which is typical for vanilla implementations of RL under partial observation in turbulent environments, the agent converges to a quasi-steady, mean-flow–oriented strategy. Using a standard SAC agent~\citep{haarnoja2018soft} \emph{without} the SSM critic and ensemble (i.e. outside REACT), both flaps settle to nearly symmetric steady deflections with small dithering about a static optimum. This reshapes the mean recirculation region and yields positive but \emph{suboptimal} drag reduction (Fig.~\ref{fig:dynamical_meanflow}c), with only limited suppression of lateral bistability and shedding (Fig.~\ref{fig:dynamical_meanflow}d). From the RL perspective, the value function in this regime encodes mostly mean-flow features, preventing the policy from learning truly dynamic, turbulent-state-dependent actions.

\emph{Ablations.} Replacing the SSM critic with MLP/LSTM, or removing the ensemble, reproducibly pushes learning toward the mean-flow–oriented regime; conversely, restoring REACT’s temporal modeling and ensembling returns the dynamics-aware policy. These ablations (see Methods~\ref{sec:critic design}), together with the open-loop replay test (see Methods~\ref{sec:closeloop_strategy_validation}), demonstrate that REACT’s algorithmic choices are necessary to discover dynamic closed-loop strategies beyond quasi-steady mean-flow modification.

\section{\textcolor{black}{Closed-loop wake control with REACT}}\label{sec:Control result}

\begin{figure}
    \centering
    \includegraphics[width=0.999\linewidth,height=0.71\textheight,keepaspectratio]{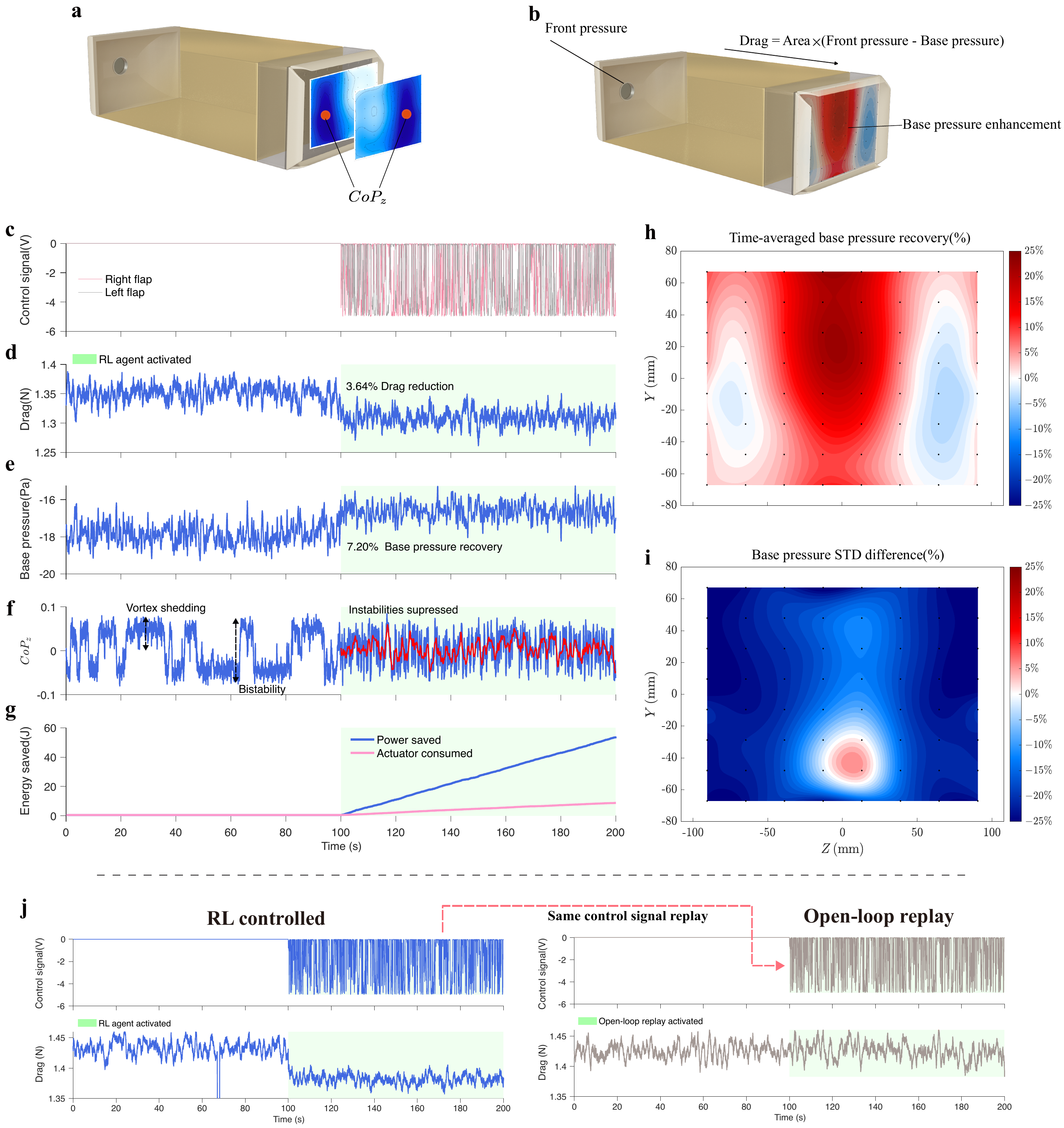}
    \caption{\textbf{The onboard control performance of the REACT system.} (\textbf{a}) Schematic illustration of spanwise ($z$-direction) center of base pressure ($CoP_z$) switching due to turbulent flow instabilities. (\textbf{b}) Schematic illustration of the relationship between base pressure and aerodynamic drag. 
 (\textbf{c}) Real-time control signals executed by the RL agent.
 (\textbf{d}) Aerodynamic drag measured by the load cell (0.6 s running mean). 
 (\textbf{e}) Spatially-averaged base pressure time series from 64 pressure sensors (0.6 s running mean). 
 (\textbf{f}) Time series of $CoP_{z}$.
 (\textbf{g}) \textcolor{black}{The cumulative energy saving and consumed.} (\textbf{h}) Contour of percentage change in base pressure. An enhancement of base pressure corresponds to a reduction in drag. (\textbf{i}) Contour of percentage change in the standard deviation of base pressure. \textcolor{black}{(\textbf{j}) Replaying the RL agent's recorded actions in open loop, starting from random initial conditions, does not reproduce the drag reduction.}}
\label{fig:timeseriessynchronise_withcontour}
\end{figure}

Within 300 wind-tunnel training episodes (40 seconds each), the REACT agent converges to a dynamical closed-loop policy that delivers net energy savings by stabilizing turbulent wake instabilities. 
At a freestream velocity of 15\,m/s ($Re =220{,}500$), policy activation at $t = 100$ s produces an immediate rise in base pressure, a reduction in drag, and a monotonic increase in cumulative energy savings (Fig.~\ref{fig:timeseriessynchronise_withcontour}(c)–(g); shaded region). These gains arise from active dynamical suppression of the dominant lateral instabilities in the wake, achieved through finite-amplitude, broadband flap motions that directly oppose the lateral unstable dynamics (Fig.~\ref{fig:timeseriessynchronise_withcontour}(c))).

To quantify lateral symmetry breaking instabilities, which dominate the wake dynamics~\cite{grandemange2013bi, brackston2016stochastic}, we track the spanwise center of pressure $CoP_z$, defined as
\begin{equation}
CoP_{z}(t) = \frac{1}{W \iint_A p(z, t) \,\mathrm{d} A} \iint_A p(z, t) \cdot z \,\mathrm{d} A,
\end{equation}
where $W$ is the rear-base width, $p(z, t)$ the rear-base pressure, and $A$ the rear-base area. The temporal evolution of $CoP_z$ is a direct proxy for wake asymmetry and oscillations (Fig.~\ref{fig:timeseriessynchronise_withcontour}(a)); base-pressure recovery correlates with drag reduction (Fig.~\ref{fig:timeseriessynchronise_withcontour}(b)).
Before control, the turbulent wake exhibits low-frequency lateral bistability~\cite{grandemange2013bi,brackston2016stochastic} superimposed with higher-frequency vortex shedding (Fig.~\ref{fig:timeseriessynchronise_withcontour}(f)). Bistability manifests itself as irregular switching in $CoP_{z}$ between two asymmetric states. Upon  activation, the policy recenters $CoP_{z}$, dynamically suppressing bistable switching and weakening the high-frequency content. This stabilization yields a  7.20\% increase in base pressure and a 3.64\% drag reduction (Fig.~\ref{fig:timeseriessynchronise_withcontour}(d)–(e)). Notably, no symmetry constraint is encoded in the algorithm; tasked solely with maximizing energy efficiency, the agent autonomously discovers wake symmetrization as the optimal strategy.

The spatial signatures of rear pressure are consistent with this mechanism. Time-averaged pressure maps show local recovery up to 25\%, concentrated near the base center (Fig.~\ref{fig:timeseriessynchronise_withcontour}(h)). 
The percentage change in the standard deviation of pressure (Fig.~\ref{fig:timeseriessynchronise_withcontour}(i)), reveals broad suppression of unsteady fluctuations across the base, corroborating the reduction of unsteady mechanisms inferred from  $CoP_{z}$ (Fig.~\ref{fig:timeseriessynchronise_withcontour}(f)).

\textcolor{black}{Benchmarking against conventional model-based controllers (Table~\ref{tab:benchmark_RL_traditional}) further highlights the advantage of the learned policy within the REACT framework. Previous linear controllers act predominantly in a higher-frequency range ($St = 0.13$--$0.22$), where they amplify vortex-shedding and shear-layer dynamics through suboptimal flap motions~\cite{brackston2016stochastic}. By contrast, the RL agent selectively excites a lower-frequency regime centred at $St = 0.02$, a strategy whose physical basis is examined in Section~\ref{sec:control mechanism}. In aerodynamic terms, the RL controller delivers greater base-pressure recovery and drag reduction, achieving nearly twice the drag reduction of the loop-shaped controller and more than four times that of the proportional controller. We also report the actuation power, expressed as a percentage of the net energy saved, together with the net energy saved over a ten-minute run at $15,\mathrm{m,s^{-1}}$ (Table~\ref{tab:benchmark_RL_traditional}). The actuation power required by the RL agent accounts of $15.8\%$ of the power saved, and, owing to its larger drag reduction, it is estimated to deliver a net energy saving approximately $2$--$3.7$ times greater than that of the model-based controllers.
}

\textcolor{black}{The drag-reduction signal serves two key purposes here. First, it provides a performance metric showing that the RL agent discovers a more effective control strategy than the reference model-based approaches~\citep{brackston2016stochastic}. Second, the open-loop replay test in Fig.~\ref{fig:timeseriessynchronise_withcontour}(j) shows that replaying the learned actuation sequence without feedback fails to reproduce either the drag reduction or the associated energy savings (see also Methods~\ref{sec:closeloop_strategy_validation}). This demonstrates that the agent operates in a state-dependent manner, responding to instantaneous flow instabilities rather than merely executing a fixed actuation pattern. Such behaviour complements existing mean-flow-oriented strategies and points to a new route for applying RL to rapid, dynamics-aware turbulence control.}

\begin{table}
\centering
\caption{\textcolor{black}{\textbf{Benchmark of REACT against model-based baselines.} Comparison of dominant actuation frequency $St_{\mathrm{max}}$, drag reduction $\Delta D$, base-pressure increase $\Delta P$, actuation power $P_{\mathrm{act}}$ expressed as a percentage of aerodynamic power saving, and net energy saved $E_{\mathrm{net}}$ over a ten-minute run.}}
\label{tab:benchmark_RL_traditional}
{\color{black}
\begin{tabular}{lccccc}
\toprule
Controller & $St_{\mathrm{max}}$ & $\Delta D$ (\%) & $\Delta P$ (\%) & $P_{\mathrm{act}}$ (\%) & $E_{\mathrm{net}}$ (J) \\
\midrule
REACT RL controller & 0.02 & \textbf{$3.64 \pm 0.09$} & \textbf{$7.20 \pm 0.26$} & 15.8 & 372 \\
Loop-shaped~\cite{brackston2016stochastic} & 0.13 & 2.00 & 3.90 & 24.0 & 185 \\
Filtered~\cite{brackston2016stochastic} & 0.19 & 1.60 & 3.70 & 15.0 & 165 \\
Proportional~\cite{brackston2016stochastic} & 0.22 & 0.90 & 1.70 & 11.0 & 108 \\
\bottomrule
\end{tabular}
}
\end{table}

\section{Mechanisms of turbulence suppression}\label{sec:control mechanism}
\begin{figure}
  \centering
  \includegraphics[width=\linewidth]{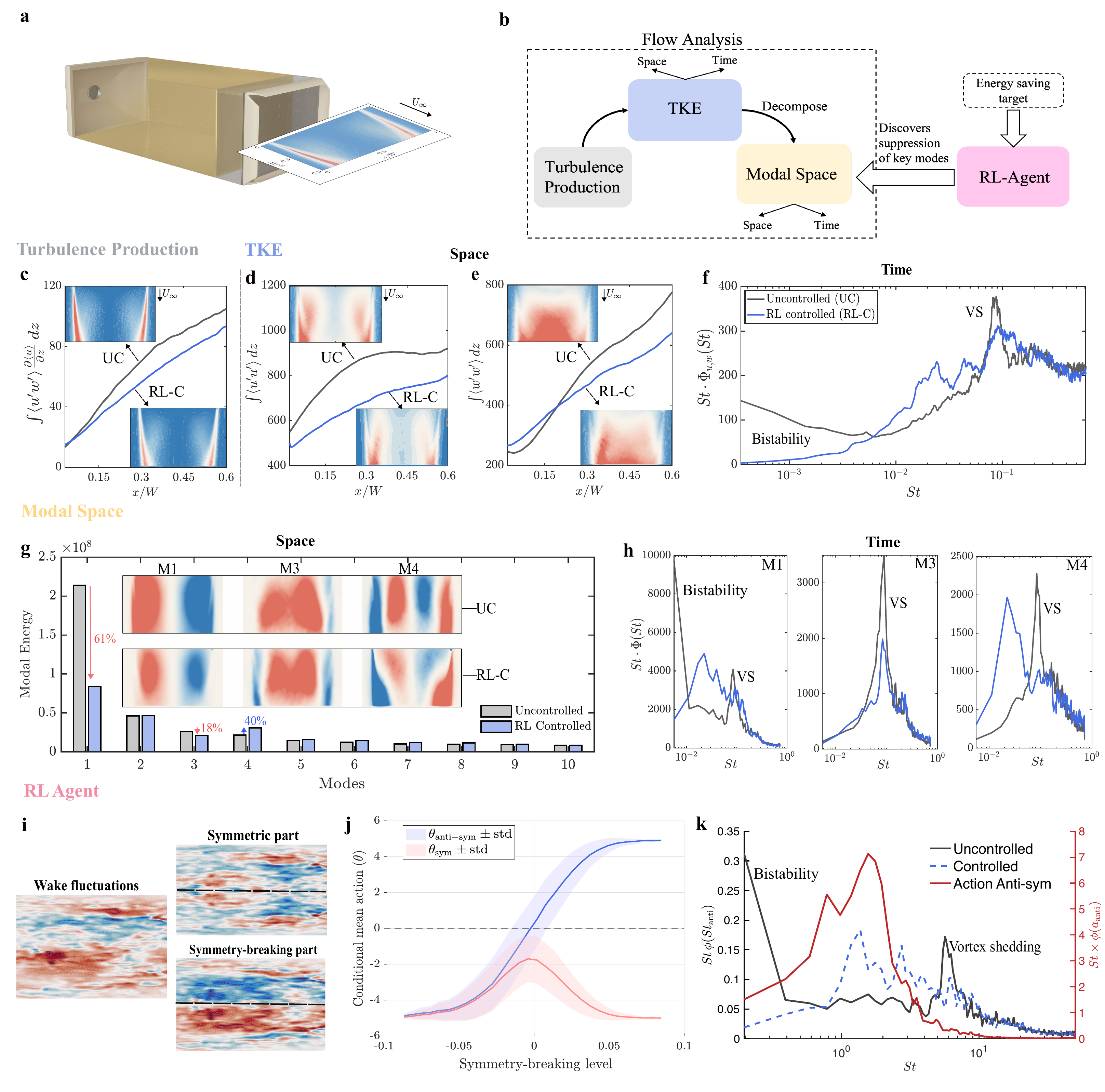}
  \caption{\textbf{Mechanism of turbulence suppression.}
  (\textbf{a}) Experimental layout and TKE-production map at $U_\infty=15$ m/s (FOV: $z/W\in[0,1.2]$, $x/W\in[0,0.6]$).
  (\textbf{b}) Analysis workflow linking TKE, production and modal space to the RL agent.
  (\textbf{c–e}) Streamwise distributions of spanwise-integrated (\textbf{c}) production, (\textbf{e}) $\langle u'u'\rangle$, and (\textbf{e}) $\langle w'w'\rangle$, with contours for uncontrolled (UC) and RL-controlled (RL-C) cases.
  (\textbf{f}) Frequency-premultiplied spectra of $u'$ (see Methods~\ref{method: Total Temporal Coefficients}).
  (\textbf{g}) Spatial POD modes M1, M3, M4 before (top) and after (bottom) control with changes in POD energy (bars).
  (\textbf{h}) Premultiplied spectra of the temporal coefficients of M1, M3, M4 showing bistability and shedding peaks.
\textcolor{black}{(\textbf{i}) Symmetry-based wake-fluctuation ($u'$) decomposition.
 (\textbf{j}) Reduced-order state--action map as a function of the symmetry-breaking level, $\Phi_{\mathrm{symb}}$.
    (\textbf{k}) Power spectral density of $\Phi_{\mathrm{symb}}$ for the uncontrolled and controlled wakes, together with the spectrum of $\theta_{\mathrm{anti\text{-}sym}}$.}}
  \label{fig:Control Mechanism}
\end{figure}
To uncover how REACT achieves drag reduction, we examine how the RL agent reshapes the turbulent wake. We quantify performance using turbulent kinetic energy (TKE, $\langle u'u' \rangle + \langle w'w' \rangle$) and turbulence production ($\langle u'w' \rangle \, \frac{\partial \langle u \rangle}{\partial z}$) from two-component planar velocity measurements. In statistical steady state, the power input required to overcome aerodynamic drag is dissipated in the flow, and is balanced by the production of TKE~\cite{pope2000turbulent}.
REACT suppresses both quantities across space and time, indicating effective mitigation of wake turbulence. To further reveal the turbulent flow dynamics targeted by the RL agent, we apply Proper Orthogonal Decomposition (POD)~\cite{Lumley1967}, which identifies coherent flow mechanisms ranked by their TKE contribution~\cite{Berkooz1993}. Without explicit physics priors, the RL agent discovers how to autonomously suppress the dominant coherent modes based solely on the energy-saving objective.

The time-averaged spatial contour of turbulence production (Fig.~\ref{fig:Control Mechanism}(c)) reveals a substantial weakening within the wake shear layer, reflecting suppressed TKE generation and reduced instability. The spanwise-integrated production curve \(\int \langle u'w' \rangle \, \frac{\partial \langle u \rangle}{\partial z} \, dz\) shows a consistent reduction across all streamwise locations under RL control (RL-C) compared to the uncontrolled case (UC), averaging 16.89\%, directly decreasing TKE. Consistently, both TKE components decrease: streamwise fluctuations \( \langle u'u' \rangle \) attenuate within the wake cores (Fig.~\ref{fig:Control Mechanism}(d)), and spanwise fluctuations \( \langle w'w' \rangle \) are reduced downstream (Fig.~\ref{fig:Control Mechanism}(e)). A localized increase in \( \langle w'w' \rangle \) near the rear surface of the vehicle is attributed to the lateral actuation of the flaps. Their integrated magnitudes fall by 19.6\% and 7.3\%, respectively.

Regarding the temporal dynamics, the RL agent learned to attenuate two different coherent instability mechanisms characteristic of bluff-body wakes, reminiscent of spatio-temporal symmetry breaking in laminar regimes \cite{grandemange2013bi,rigas2014low,brackston2016stochastic,callaham2022empirical}, appearing as peaks in the TKE spectrum (Fig.~\ref{fig:Control Mechanism}(f)). The low-frequency peak associated with bistable wake switching vanishes entirely under control (Fig.~\ref{fig:Control Mechanism}(f)), demonstrating active stabilization of stochastic asymmetry. High-frequency vortex shedding is partially suppressed, consistent with reductions in \( CoP_z \) oscillations. Notably, energy increases at intermediate frequencies, suggesting that REACT deliberately excites less drag-sensitive dynamics as a by-product of flap motion.

The orthogonal modal decomposition of the wake further clarifies the discovered control strategy. The dominant mode M1, which couples bistable switching with vortex shedding, loses 61\% of its energy under control (Fig.~\ref{fig:Control Mechanism}(g)). Its spectrum shows complete suppression of the bistable dynamics and partial suppression of vortex shedding (Fig.~\ref{fig:Control Mechanism}(h)). Mode M2, representing symmetric wake fluctuations~\cite{berger1990coherent, rigas2014low} , remains largely unaffected, consistent with its minimal role in drag. Mode M3, linked to higher-frequency shedding, is reduced by 18\%. In contrast, mode M4 is not suppressed but modulated: its spectral peak shifts from vortex-shedding frequencies to a new component at \( St = 0.02 \), matching the flap actuation. Although this shift amplifies fluctuations near the base edges, it relocates energy into dynamics with negligible drag impact. Unlike conventional controllers, which often exacerbate vortex shedding while only partly suppressing bistability, REACT uncovers autonomously a control strategy by selectively suppressing or redirecting coherent modes to achieve net drag reduction.

This multi-modal suppression distinguishes REACT from prior model-based feedback strategies, which face an inherent trade-off: suppressing one instability often amplifies others. Opposition control, as demonstrated by Li et al.~\citep{li2016feedback}, successfully symmetrizes the bimodal wake but enhances antisymmetric vortex shedding, yielding only 2\% base-pressure recovery despite effective bistability suppression. Similarly, linear controllers targeting bistability amplify higher-frequency shear-layer instabilities ($St = 0.13$--$0.22$), negating drag benefits~\citep{brackston2016stochastic}, while model-based control approaches failed to suppress vortex shedding even independently in 3D wakes \citep{brackston2018modelling}. REACT circumvents this trade-off: the learned policy selectively suppresses bistability and vortex shedding while redirecting residual energy into dynamically benign frequencies ($St = 0.02$), achieving net drag reduction without the adverse mode amplification that limits model-based designs. Unlike model-based approaches that require explicit specification of target modes and control gains, REACT autonomously discovers jointly optimal suppression across the multi-modal instability landscape.

\textcolor{black}{The strategy discovered by REACT is state-dependent to the turbulent fluctuations. Since both wake bistability and vortex shedding are manifestations of symmetry-breaking dynamics and are broadly observed in bluff-body wakes \citep{cadot2026review}, it is natural to decompose the wake field into symmetric and antisymmetric components in order to isolate the dynamically relevant subspace for interpreting the learned control law. We decompose the RL actions into symmetric and anti-symmetric components, defined as:}

\begin{equation}
\color{black}
\begin{aligned}
\theta_{\text{anti-sym}} &= \theta_{\mathrm{left}} - \theta_{\mathrm{right}}, \\
\theta_{\text{sym}}      &= \theta_{\mathrm{left}} + \theta_{\mathrm{right}}.
\end{aligned}
\end{equation}

{\color{black}The wake fluctuations($u'$) are likewise decomposed into symmetric and anti-symmetric components, as illustrated in Fig.~\ref{fig:Control Mechanism}(i), according to:
\begin{equation}
\begin{aligned}
u'_{\mathrm{sym}}(x,z,t)
&=
\frac{1}{2}\left[
u'(x,z,t)+u'(x,-z,t)
\right],
\\
u'_{\mathrm{anti\text{-}sym}}(x,z,t)
&=
\frac{1}{2}\left[
u'(x,z,t)-u'(x,-z,t)
\right].
\end{aligned}
\end{equation}}

{\color{black}To visualize the structure of the learned feedback law, Fig.~\ref{fig:Control Mechanism}(j) shows a statistical reduced-order state--action map in the symmetry-decomposed space. The antisymmetric component of the actuation($\theta_{\text{anti-sym}}$) exhibits a nonlinear dependence on the wake symmetry-breaking level, defined as
\begin{equation}
\Phi_{\mathrm{symb}}
=
\int_{\Omega} u'_{\mathrm{anti\text{-}sym}}(x,z)\,\mathrm{d}x\,\mathrm{d}z .
\end{equation}
This result suggests that the controller does not respond arbitrarily, but instead acts in a structured manner on the symmetry-breaking wake dynamics associated with bistability and vortex shedding.

To further examine this mechanism, Fig.~\ref{fig:Control Mechanism}(k) compares the power spectral density of $\Phi_{\mathrm{symb}}$ in the uncontrolled and controlled cases, together with that of $\theta_{\mathrm{anti\text{-}sym}}$. Under control, the total spectral energy of $\Phi_{\mathrm{symb}}$, quantified by the area under the spectrum, is reduced. In particular, both the low-frequency bistable dynamics and the higher-frequency vortex-shedding component are suppressed. Moreover, the frequency band amplified by the controller aligns closely with the dominant band of $\theta_{\mathrm{anti\text{-}sym}}$, supporting the interpretation that the learned actuation is selectively organized to regulate anti-symmetric wake motions across the relevant dynamical scales.

We further notice that the RL agent discovers a mean-flow regulation component in the symmetric sub-space(Fig.~\ref{fig:Control Mechanism}(j)). The symmetric component of the actuation, $\theta_{\mathrm{sym}}$, is more consistent with mean-flow shaping and the overall actuation level. For example, near $\Phi_{\mathrm{symb}}=0$, a non-zero value of $\theta_{\mathrm{sym}}$ is still present. In this regime, when $\langle|\theta_{\mathrm{anti\text{-}sym}}|\rangle\approx 0$ and $\theta_{\mathrm{sym}}<0$, both flaps tend to move slightly inwards in a nearly symmetric manner. By contrast, at larger values of $\Phi_{\mathrm{symb}}$, the actuation becomes predominantly antisymmetric, with $|\theta_{\mathrm{sym}}|\approx |\theta_{\mathrm{anti\text{-}sym}}|$ in magnitude. This suggests that the controller redistributes its actuation effort according to the instantaneous wake state: symmetric input is used when mean-flow adjustment is beneficial, whereas antisymmetric input dominates when direct regulation of wake asymmetry is required. }

Attenuation of the coherent structures requires actuation in the production region. The REACT system operates within this production region and acts primarily on 
$u^{\prime}$, i.e. on the rapid, and multi-scale large-scale flow instabilities, 
which necessitates real-time, state-dependent control. On the other hand, mean-flow--oriented strategies primarily reshape the mean flow by modifying \(\left\langle u \right\rangle\) 
or the mean shear \(\frac{\partial \left\langle u \right\rangle}{\partial z}\), 
altering the structure and intensity of the production region in a quasi-steady manner. Finally, because the REACT targets dynamics in the production region rather than the small dissipative scales, a 100\,Hz sensing--actuation loop is sufficient.

\section{\textcolor{black}{Extrapolation across Reynolds number}}

\begin{figure}
    \centering
     \includegraphics[width=1\linewidth]{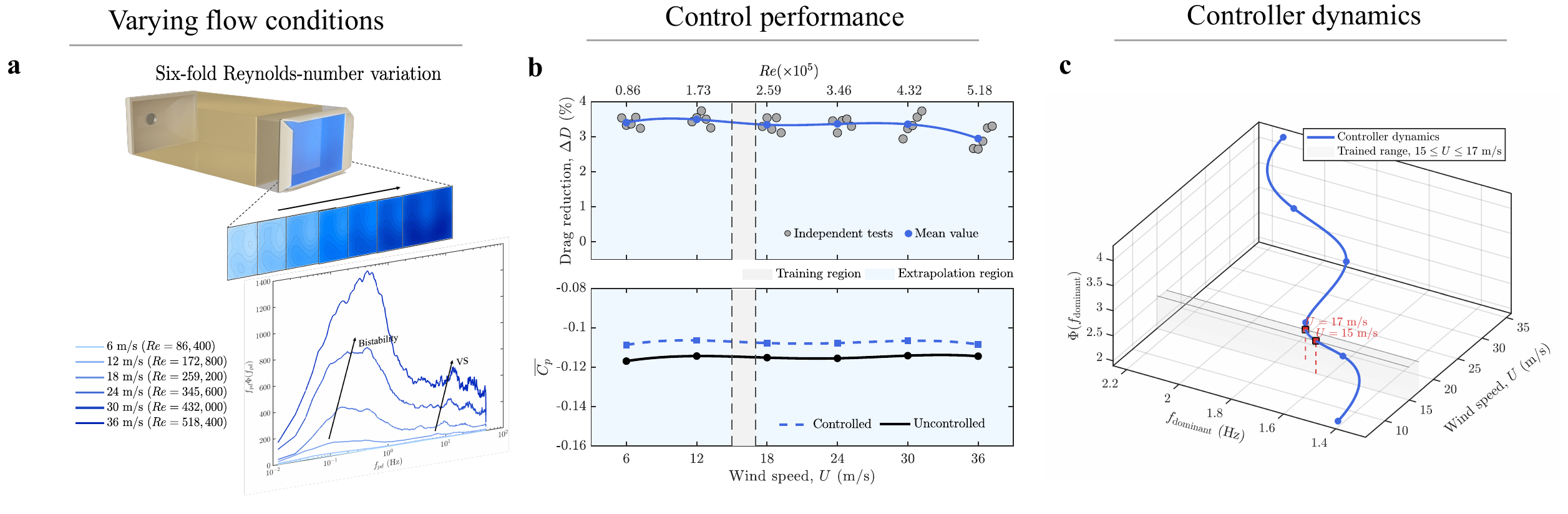}
\caption{\textbf{\textcolor{black}{Extrapolation performance of REACT.}}
(\textcolor{black}{\textbf{a}) Increasing wind speed, and hence Reynolds number, simultaneously increases the mean base-pressure magnitude, the dominant instability frequencies, and the fluctuation intensity of the wake. These changes are shown by the spatial distributions of the time-averaged base pressure and the frequency-premultiplied spectra of the lateral pressure differential across different wind speeds.
(\textbf{b}) Time-averaged drag reduction and base-pressure enhancement achieved by the generalized controller across test cases spanning a sixfold variation in wind speed. Each wind speed includes five independent runs. The trained region is indicated by the grey shaded area, and the extrapolation regions are indicated by the blue shaded areas.
(\textbf{c}) Evolution of the control trajectory in the space spanned by wind speed, the controller's dominant frequency \(f_{\mathrm{dominant}}\), and the corresponding power spectral density of the actuator signal, \(\Phi(f_{\mathrm{dominant}})\). The trained region is indicated by the grey shaded area.
}}
    \label{fig:Controller Trajectory}
\end{figure}

\textcolor{black}{Discovering transferable policy has become a central challenge in reinforcement-learning-based flow control, particularly for transferring learned policies across flow configurations \citep{wang2026physics, chatzimanolakis2024learning}. In this section, we assess the robustness of REACT under varying flow conditions. Despite being trained offline at only two freestream velocities, 15 and 17\,m/s, REACT maintains effective drag reduction across a sixfold velocity range, from 6 to 36\,m/s, corresponding to Reynolds numbers from \(86{,}400\) to \(518{,}400\). This extrapolation across vehicle speeds is challenging because changes in freestream velocity alter: (i) the frequency and amplitude of the dominant wake instabilities (Fig.~\ref{fig:Controller Trajectory}(a)); (ii) the aerodynamic loading and mean base pressure (Fig.~\ref{fig:Controller Trajectory}(a)); and (iii) the effective control authority of the rear flaps. REACT adapts autonomously to these coupled changes, sustaining effective performance under flow conditions not observed during training.}

\textcolor{black}{REACT achieves extrapolation in two complementary ways. First, a physics-consistent scaling formulation ensures amplitude invariance with respect to $Re$ changes for the actor and critic inputs: the training is performed in dimensionless space independent of $Re$ (pressure coefficients $C_p$, drag coefficient $C_d$), so that input magnitudes remain approximately invariant across speeds at turbulent regimes.  
In general, invariance of the state-feedback mapping to parameter changes facilitate extrapolation.
Second, since instability frequencies shift with $Re$, the actor (policy function) and critic (value function) are conditioned explicitly on Reynolds number during offline training, enabling temporal adaptation when residual speed-dependent effects (e.g. variation in flap authority or frequency scaling) become relevant (see also Methods~\ref{sec:Methods_generalization}). }

\textcolor{black}{Figure~\ref{fig:Controller Trajectory}(b) shows the trends in drag reduction and base-pressure enhancement, with the training and extrapolation regions indicated by different shaded areas. The results demonstrate that the controller remains robust across the sixfold Reynolds-number range tested. Although a slight reduction in performance is observed at the highest wind speed, effective drag reduction is still sustained. The corresponding base-pressure enhancement exhibits a trend consistent with the drag-reduction performance. Importantly, this robustness does not arise from simply reusing a fixed control strategy across all flow conditions. Rather, the agent adapts to changes in wind speed by varying both the dominant actuation frequency and the actuation strength, as shown in Fig.~\ref{fig:Controller Trajectory}(c). The training range, indicated by the shaded volume, occupies only a narrow region of the full trajectory space, emphasizing the controller's ability to extrapolate beyond the conditions encountered during training.} \textcolor{black}{These results show that the physics-consistent formulation, offline training, and Reynolds-conditioned learning enable a single RL controller to remain effective across sixfold wind speed variation.}

\section{Discussion}

REACT achieves both dynamical turbulence suppression and net energy savings through direct policy training in a high-Reynolds-number wind-tunnel environment. This is enabled by several advances. First, a sequential and SSM-based ensemble RL architecture provides the expressiveness needed to efficiently and robustly learn under partially observed, high-dimensional and turbulent dynamics. Second, a physics-consistent training space combined with parametric conditioning and offline training enables robust generalization across flow regimes. Third, real-time learning and control are enabled by a low-latency pipeline connecting GPU-based edge computation with on-vehicle perception and actuation. \textcolor{black}{Together, they provide a framework for autonomous, state-dependent and interpretable control of turbulent wake dynamics in physical experimental setting.}

In turbulent bluff-body wakes, unsteady spatio-temporal coherent structures channel energy from the mean flow into turbulence. A substantial body of model-free applications, including RL-based studies, has shown that strategies acting through mean-flow modification can deliver significant drag reduction. Consistent with this, we find that vanilla RL agents in our partially observable turbulent environment can converge to quasi-steady policies in which the flaps make small adjustments about a static optimum; these solutions reduce drag but attenuate little the instantaneous lateral instabilities in the  turbulent wake. Therefore, we instead exploit the ability of RL to discover a dynamic, state-dependent policy in complex turbulent environments. The resulting flapping motion actively counters instantaneous wake asymmetries and suppresses multi-modal coherent structures (bistability and vortex shedding), rather than a single coherent structure alone~\cite{brackston2016stochastic, pastoor2008feedback}. Such feedback policies complement mean-flow–modification approaches by providing a dynamics-aware, but model-free direction to control of turbulent dynamics.

\textcolor{black}{We also acknowledge several limitations of the present approach. The study focuses on a canonical bluff-body geometry in a wind-tunnel environment, which allows the key mechanisms by which RL-based feedback modifies turbulent wake dynamics to be isolated clearly. More complex body shapes and geometries remain to be explored in future work. The present study is not performed in a full-scale or industrial-scale vehicle deployment. But we anticipate that similar wake structure will appear in the industrial-vehicle (See Methods~\ref{methods:current model}). In addition, we employ a deliberately minimal actuation and sensing layout, consisting of two rear flaps driven by base-pressure measurements. This choice reflects realistic constraints on actuator count and sensor availability in automotive applications. More complex configurations, including distributed actuation or additional on-body and off-body sensing, could further improve control performance and robustness.}

Beyond these scope choices, REACT advances RL from simulation to \emph{hardware-in-the-loop} operation in turbulent flows, achieving dynamics-aware control within latency limits and energy budgets. \textcolor{black}{More broadly, these results suggest the potential of learning-based model-free controllers for dynamics-aware operation in turbulent flow control applications.}

\newpage
\section{Methods}\label{sec11}

\subsection{Wind-tunnel environment and current model}\label{methods:current model}

The experiments were carried out in the temperature-stabilized T2 wind-tunnel (Fig.~\ref{fig:windtunnelandmodel}(a)) at Imperial College London, which has a test section (Fig.~\ref{fig:windtunnelandmodel}(b)) measuring 1.11\,m in height, 1.66\,m in width, and 4\,m in length. The blockage ratio of the current setup is 1. 88\%, which remains well below the typical threshold to avoid significant wall interference. The freestream velocity is regulated by a proportional--integral--derivative (PID) controller, maintaining the desired flow speed with an accuracy of 0.25\%.

\begin{figure}
    \centering
    \includegraphics[width=\linewidth]{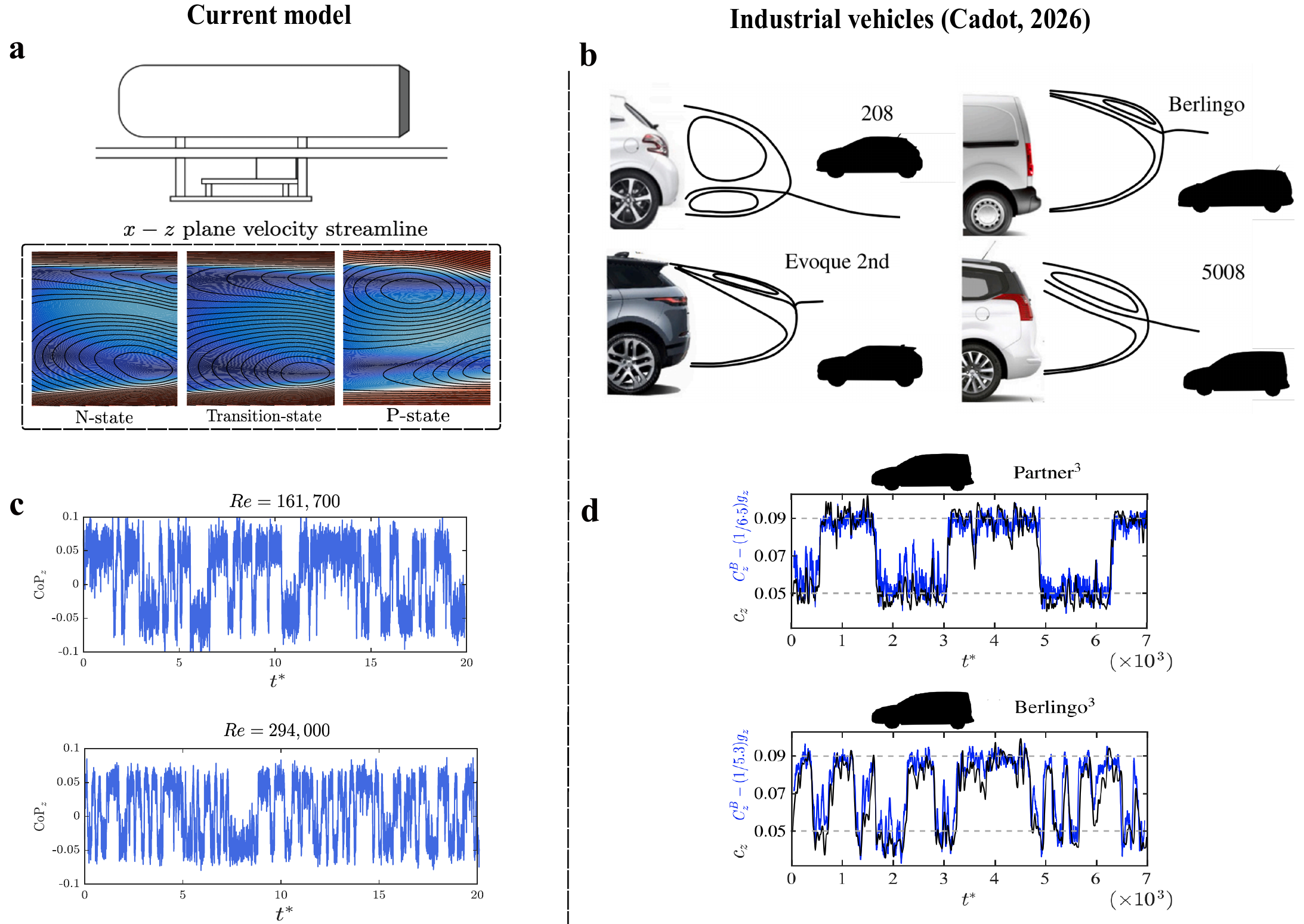}
    \caption{{\color{black}(\textbf{a}) Asymmetric turbulent wake captured by the present Ahmed model.
(\textbf{b}) Asymmetric turbulent wake observed in industrial vehicles.
(\textbf{c}) Bistable switching and vortex shedding dynamics captured by the present model, with increasing Reynolds number corresponding to different dominant switching frequencies.
(\textbf{d}) Bistable switching and vortex shedding dynamics observed in industrial vehicle showing distinct switching behaviours.}}
    \label{fig:dynamics_capturing}
\end{figure}

{ \bf Ahmed-body model.} {\color{black}The square-back Ahmed body is adopted here as a canonical system that isolates the same wake physics observed in a growing class of road-vehicle wakes: the global symmetry-breaking instability of the separated recirculating flow that gives rise to large-scale asymmetric states, discrete wake reversals and bistable switching between them. This mechanism is now well established for square-back bluff bodies and has also been identified in realistic automotive configurations. In full-scale studies of production vehicles, discontinuous transitions between opposite wake states and associated bistable dynamics have been reported for the industrial vehicle with the observed behaviour explicitly interpreted as the same global 
 instability as that of the square-back Ahmed body \citep{bonnavion2019asymmetry}. More broadly, a recent review concluded that asymmetric recirculating wakes, wake reversal and bistability constitute a consistent pattern across industrial ground vehicles, with the underlying property being the same steady instability as in the Ahmed-body wake \citep{cadot2026review}. Earlier measurements on the industrial vehicle likewise linked real-vehicle bistability to this wake-instability mechanism rather than to a geometry-specific artefact \citep{bonnavion2017multistabilities}. Recent experiments on an industrial model further reinforce this interpretation by showing a sharp transition between two asymmetric wake states, \(P\) and \(N\), with bistable dynamics and substantial related drag variations \citep{keirsbulck2024underbody}. These results support the view that the Ahmed body as current model is a scientifically meaningful testbed for studying control laws that target state selection and instability-mediated wake dynamics in realistic road-vehicle flows.

As shown in Fig.~\ref{fig:dynamics_capturing}\textbf{(a--b)}, we compare the asymmetric wake captured by the present model with the asymmetric wakes reported for production vehicles in previous studies. The wake topology is qualitatively similar, although the strength and detailed form of the asymmetry depend on the specific vehicle geometry. We further compare the time-resolved dynamics of wake asymmetry, characterised here by the wake centre of pressure (\(CoP_z\)), in Fig.~\ref{fig:dynamics_capturing}\textbf{(c)} with those reported for an industrial vehicle in Fig.~\ref{fig:dynamics_capturing}\textbf{(d)}. In both cases, the dynamics are characterised by bistable switching coupled with oscillations associated with shedding instabilities. Moreover, changes in vehicle geometry (for example, from the Peugeot Partner to the Citro\"en Berlingo) lead to shifts in the dominant dynamical frequency; a similar trend is captured in the present experiments when the Reynolds number is varied, as shown in Fig.~\ref{fig:dynamics_capturing}\textbf{(c)}.
The Ahmed-body configuration therefore provides a controlled yet physically relevant platform for isolating and interrogating wake mechanisms that are also present in industrial vehicle aerodynamics.}

The length of the model is 0.6m. The base measures 0.216m $\times$ 0.160\,m and is elevated 0.028\,m above a raised floor to minimize boundary layer effects while preserving realistic ground conditions (Fig.~\ref{fig:windtunnelandmodel}(f)). The same setup has been used in the previously reported control experiments in \cite{brackston2016stochastic}. 

{ \bf Sensors.} A force balance (ATI Gamma-IP68 load cell) located beneath the raised floor, outside the flow path, connects the model to the tunnel and enables accurate measurement of aerodynamic forces.
\begin{figure}\label{fig:windtunnelandmodel}
    \centering
    \includegraphics[width=0.9\linewidth]{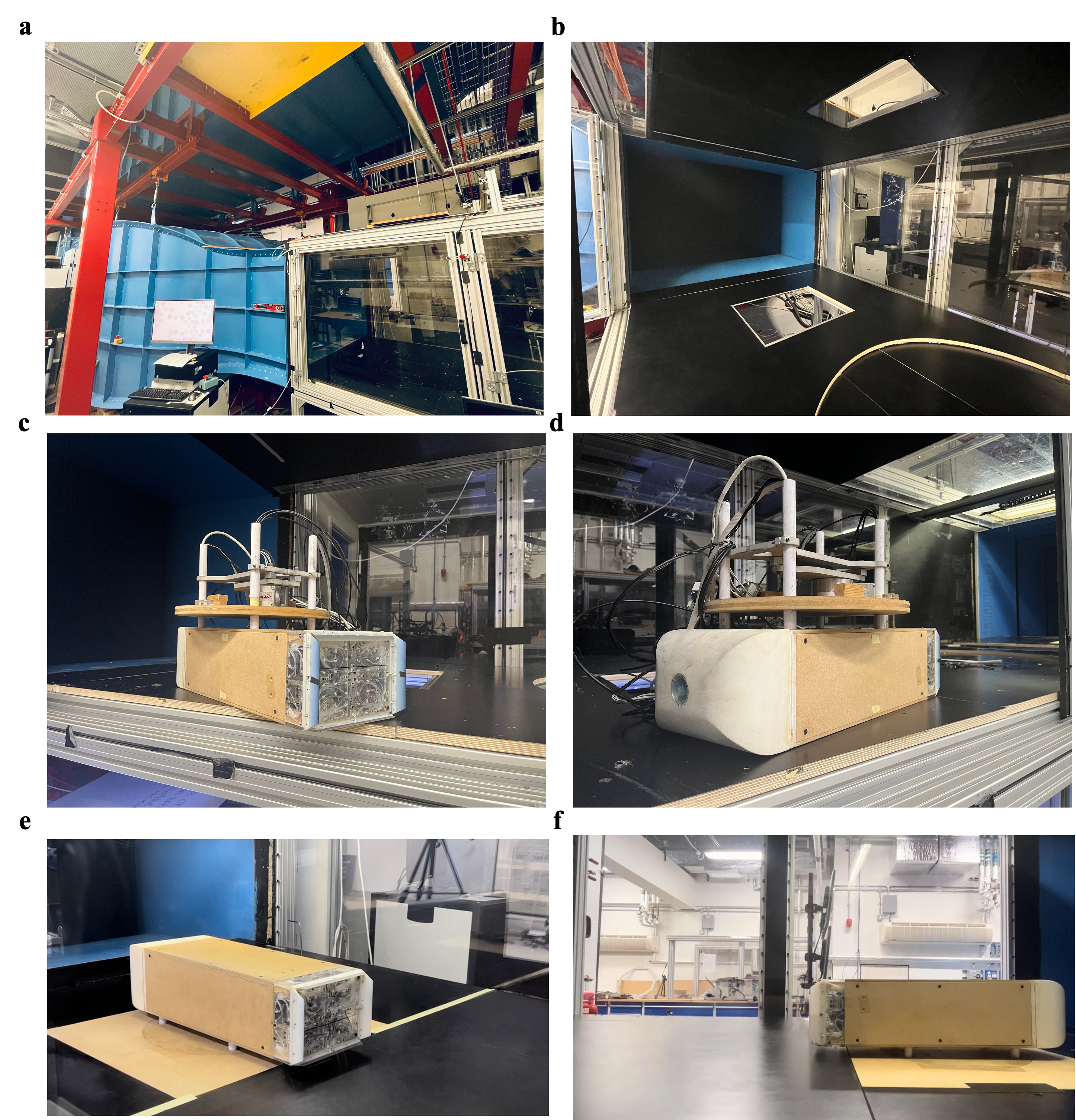}
    \caption{\textbf{The T2 wind-tunnel used and the Ahmed body model.}    (\textbf{a}) The T2 wind-tunnel facility at Imperial College London, used for the present experiments.  (\textbf{b}) The test section where the model is mounted.  (\textbf{c}) Rear view of the Ahmed body, showing integrated sensors and actuators.  (\textbf{d}) Front view of the Ahmed body.  (\textbf{e-f}) The Ahmed body during experimental deployment.}
    \label{fig:windtunnelandmodel}
\end{figure}
Rear surface pressure is monitored via 64 static pressure taps distributed across the model's base, connected to an ESP-DTC pressure scanner (Chell $\mu$DAQ2-64DTC) that streams real-time measurements to the control loop via UDP. 

{ \bf Actuators.} Wake forcing is provided by two flaps mounted at the rear side edges, spanning the full height of the body and 0.019 m in chord length.  Each flap is hinged and actuated by internal motors powered through amplifiers, with passive restoring force provided by internal springs. The power consumption is monitored through real-time measurements of motor supply voltage and current.

\subsection{Policy network}\label{sec:netword architectrue}

The policy network combines feedforward and recurrent pathways. The input to the network consists of a state vector and action in one previous step, which are processed in two parallel branches. The feedforward branch maps the current state through a linear layer with 512 hidden units followed by a ReLU activation. The recurrent branch concatenates the state and previous action, projecting them into a 512-dimensional space and feeding them into a single-layer LSTM with 512 hidden units. The outputs from both branches are concatenated, forming a 1024-dimensional feature vector, which is further transformed by two fully connected layers with 512 hidden units each. The final layers produce the mean and log standard deviation of a Gaussian policy. Actions are sampled using the reparameterization trick and mapped to the bounded action space via a $\tanh$ transformation~\cite{haarnoja2018soft}.

\subsection{Critic network}\label{sec:critic design}

\begin{figure}
    \centering
    \includegraphics[width=0.75\linewidth]{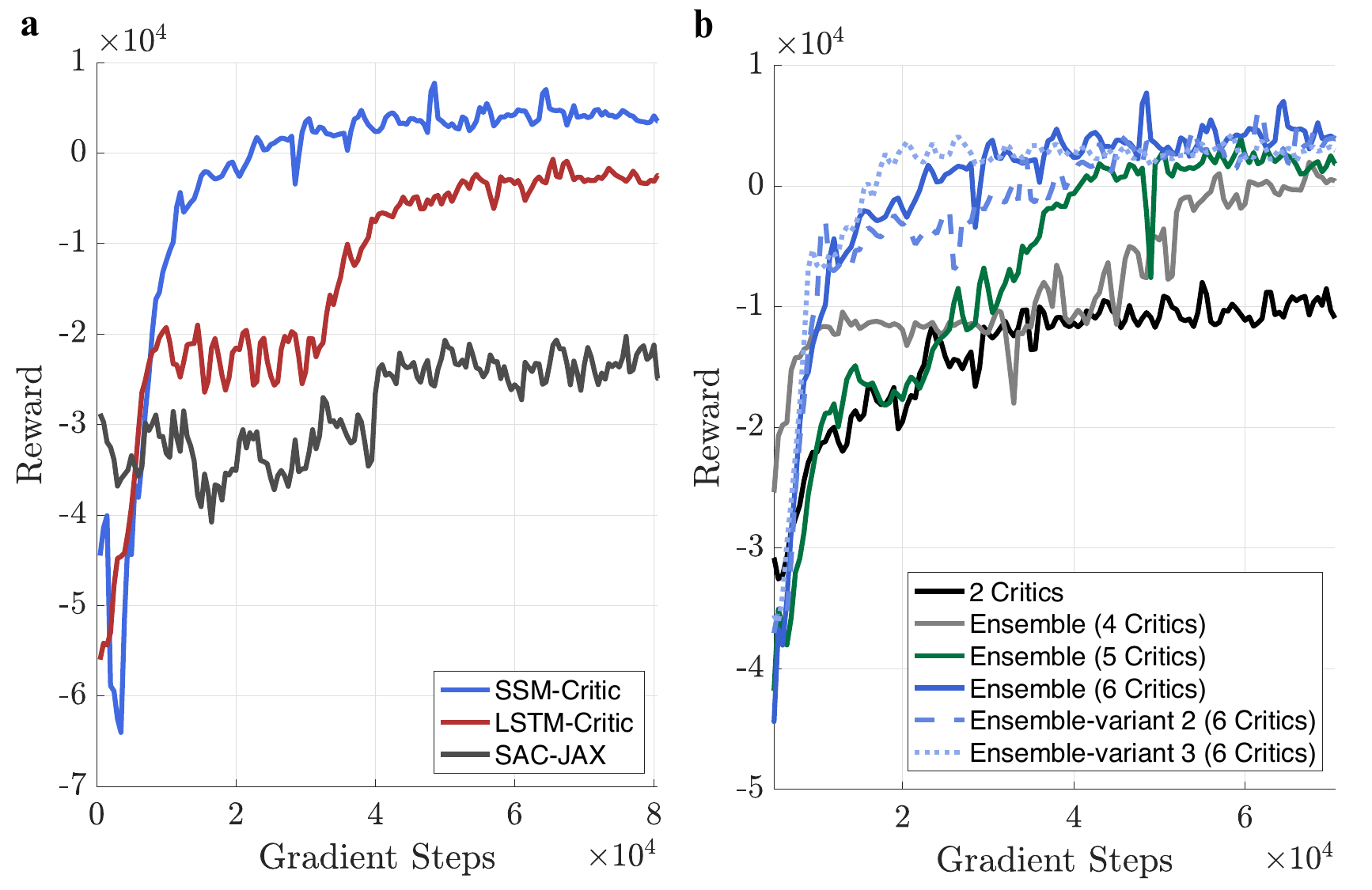}
    \caption{\textbf{The ablation study of critic type and ensemble size.}   (\textbf{a}) Comparison of reward performance among SSM-Critic, LSTM-Critic, and SAC with MLP-based actor and critic. (\textbf{b}) Reward comparison across different sizes and implementations of critic ensembles.}
    \label{fig:rewardplot}
\end{figure}

The critic design enables REACT to learn under partial observability and stochasticity typical of turbulent flows. Standard critic architectures (MLPs, LSTMs) fail to converge to dynamics-aware policies in this setting, as we demonstrate empirically below.
We address this through two coupled innovations. First, we introduce a selective state-space model (SSM) critic based on the Mamba architecture~\cite{gu2023mamba, dao2024transformers}, embedded in a sequence-to-sequence evaluator (§8.3.1–8.3.2). Unlike recurrent critics that treat temporal credit assignment uniformly, the SSM's selective gating learns to weight information flow adaptively, which proves essential for value estimation in chaotic systems where relevant timescales span orders of magnitude. Second, we aggregate multiple independently initialized critics using conservative min-ensemble target computation (§8.3.3) to suppress the overestimation bias that otherwise destabilizes learning under the high stochasticity of turbulent observations.
Crucially, neither component alone is sufficient: SSM critics without ensembling exhibit unstable training, while ensembles of MLP or LSTM critics converge to quasi-steady policies that fail to discover state-dependent control (Fig.~\ref{fig:rewardplot}).  We benchmark the SSM critic against LSTM and MLP alternatives and ablate ensemble size in §8.3.3, demonstrating that both the architecture choice and ensemble design are necessary for REACT's performance.

{\color{black} \subsubsection{Selective state–space block (SSM).}

Given input $u_t\!\in\!\mathbb{R}^{D_{\mathrm{in}}}$, state $x_t\!\in\!\mathbb{R}^{N}$, and output $y_t\!\in\!\mathbb{R}^{D_{\mathrm{out}}}$,
let $A\!\in\!\mathbb{R}^{N\times N}$ be a learned,  time-invariant matrix. 
A token-wise selector $s:\mathbb{R}^{D_{\mathrm{in}}}\!\to\!\mathbb{R}_{+}\times\mathbb{R}^{N\times D_{\mathrm{in}}}\times\mathbb{R}^{D_{\mathrm{out}}\times N}$ produces
\begin{equation}
(\Delta_t,B_t,C_t)=s(u_t).
\end{equation}
With $\phi(Z)=(e^{Z}-I)Z^{-1}$, we define the discretized parameters
\begin{equation}
\bar A_t=e^{\Delta_t A},
\qquad
\bar B_t=\phi(\Delta_t A)\,\Delta_t B_t .
\end{equation}
The time-varying recurrence (selective scan) is
\begin{equation}
x_t=\bar A_t x_{t-1}+\bar B_t u_t,\qquad y_t=C_t x_t,\qquad x_0=0.
\end{equation}
Equivalently, the causal input–output map can be written as
\begin{equation}
y_t=\sum_{k=1}^{t} C_t\!\left(\prod_{i=k+1}^{t}\bar A_i\right)\bar B_k\,u_k.
\end{equation}
Because $(\Delta_k,B_k,C_k)$ depend on $u_k$, this mapping is nonlinear in the input sequence and is \emph{not} a global linear convolution; it reduces to a linear convolution only when $(\Delta_t,B_t,C_t)$ are constant in $t$.}

The SSM block in our critic architecture has been benchmarked against a conventional LSTM module. Fig.~\ref{fig:rewardplot}(a) shows the reward trend of RL agents trained with each critic variant, demonstrating that the SSM-based critic outperforms its LSTM counterpart in both convergence speed and final performance. For baseline comparison, we also compare against a standard SAC implementation employing purely feedforward MLPs in both actor and critic networks (denoted as SAC-JAX), which highlights the necessity of incorporating memory-preserving structures in partially observable environments. This SAC baseline is implemented using JAX~\cite{jax2018github} within a benchmarked RL library~\cite{stable-baselines3}, indicating that a standard SAC implementation is unable to learn an effective control strategy in this turbulent environment. For further details on the Mamba architecture used, refer to ~\cite{gu2023mamba, dao2024transformers}.

\subsubsection{Sequential value evaluation.}
We embed the SSM block inside the critic alongside time-local MLP layers. Because these pointwise MLPs act independently at each step, the overall critic remains a causal sequence-to-sequence map
\[
\mathcal{F}:\ \mathbf{S}\in\mathbb{R}^{\mathcal{B}\times L\times D_{\mathrm{in}}}
\ \longmapsto\ 
\mathbf{V}\in\mathbb{R}^{\mathcal{B}\times L\times 1},
\]
%
where $\mathcal{B}$ is the batch size and $L$ the sequence length. The value at time $\tau$, $V_\tau$, depends only on the prefix $\{u_t\}_{t=1}^{\tau}$ via the causal SSM recursion, thus aggregating the past while remaining strictly causal. The latent $x_\tau$ functions as a learned low-dimensional sufficient statistic of the history in the HMM/POMDP sense~\cite{bertsekas2012dynamic}.

\subsubsection{\textcolor{black}{Ensemble critics}}

{\color{black}We maintain an ensemble of $N$ critics $\{Q_{\phi_i}\}_{i=1}^N$ with target copies $\{Q_{\tilde{\phi}_i}\}_{i=1}^N$. Each critic follows the SSM-based sequential value evaluation described above and is conditioned on $(s_t,a_t,a_{t-1})$. Let $\alpha>0$ denote the temperature, $\gamma\in[0,1)$ the discount factor, and $d_t\in\{0,1\}$ the terminal indicator. With $a'_{t+1}\sim\pi_\theta(\cdot\mid s_{t+1})$, our default target uses the minimum over the full target ensemble:
\begin{equation}\label{eq:min_aggregate}
y_t
=
r_t
+
\gamma(1-d_t)
\Big[
\min_{i=1,\dots,N} Q_{\tilde{\phi}_i}(s_{t+1},a'_{t+1})
-
\alpha \log \pi_\theta(a'_{t+1}\mid s_{t+1})
\Big].
\end{equation}
The critic ensemble is trained by the mean squared Bellman error,
\begin{equation}\label{eq:aggregate_loss}
\mathcal{L}_Q\big(\{\phi_i\}\big)
=
\frac{1}{N}\sum_{i=1}^N
\mathbb{E}\!\left[
\big(Q_{\phi_i}(s_t,a_t)-y_t\big)^2
\right],
\end{equation}
while the policy is updated against the current-ensemble minimum,
\begin{equation}
\mathcal{L}_\pi(\theta)
=
\mathbb{E}\!\left[
\alpha \log \pi_\theta(a_t\mid s_t)
-
\min_i Q_{\phi_i}(s_t,a_t)
\right].
\end{equation}

Ensemble critics are widely used in off-policy RL to improve stability and sample efficiency \cite{chen2021randomized,hiraoka2021dropout,bhatt2019crossq}. In our setting, they are particularly beneficial because turbulent wakes are stochastic and intermittently switch between metastable regimes, so trajectories rapidly decorrelate even under similar conditions. The critic ensemble therefore serves as a set of diverse value realizations, while the minimum aggregation in Eq.~\eqref{eq:min_aggregate} acts as a conservative estimate that suppresses overestimation spikes and improves robustness under partial observability and intermittent dynamics.

Different ensemble implementations are validated in the current  environment. Following the REDQ-style idea of maintaining a large ensemble and sampling only a subset for target construction \cite{chen2021randomized}, we tested a variant in which a subset $\mathcal{M}$ of size $M$ is sampled from the $N$ target critics and the Bellman target becomes
\begin{equation}
y_t
=
r_t
+
\gamma(1-d_t)
\Big[
\min_{i\in\mathcal{M}} Q_{\tilde{\phi}_i}(s_{t+1},a'_{t+1})
-
\alpha \log \pi_\theta(a'_{t+1}\mid s_{t+1})
\Big].
\end{equation}
We further considered a pessimism-weighted variant, in which critics predicting lower target values than the ensemble mean are sampled with higher probability. Specifically, defining
\begin{equation}
\bar Q(s',a')=\frac{1}{N}\sum_{j=1}^N Q_{\bar\phi_j}(s',a'),
\qquad
u_i(s',a')=\big[\bar Q(s',a')-Q_{\bar\phi_i}(s',a')\big]_+,
\end{equation}
the sampling probabilities are given by
\begin{equation}
p_i(s',a')
=
\frac{\exp\!\big(\beta u_i(s',a')\big)}
{\sum_{j=1}^N \exp\!\big(\beta u_j(s',a')\big)},
\qquad i=1,\dots,N,
\end{equation}
\begin{equation}
\mathcal{M}\sim \operatorname{WSWOR}(p_1,\dots,p_N;M),
\end{equation}
where $[x]_+ = \max(x,0)$, $\beta\ge 0$ controls the strength of the pessimistic bias, and WSWOR indicates weighted sampling without replacement.

As shown in Fig.~\ref{fig:rewardplot}b, larger critic ensembles improve training stability and generally lead to higher-return policies, although with diminishing returns at larger $N$. The alternative subset-based variants (variant 2 and variant 3) showed comparable final performance, with the uncertainty-weighted version exhibiting slightly better early-stage sample efficiency. We nevertheless adopt the full minimum-aggregation form in Eq.~\eqref{eq:min_aggregate} as the default design because it provided the most consistent and reproducible behaviour across runs.}

\subsection{\textcolor{black}{Comparison with model-based controller}}

\begin{figure}
    \centering
    \includegraphics[width=\linewidth]{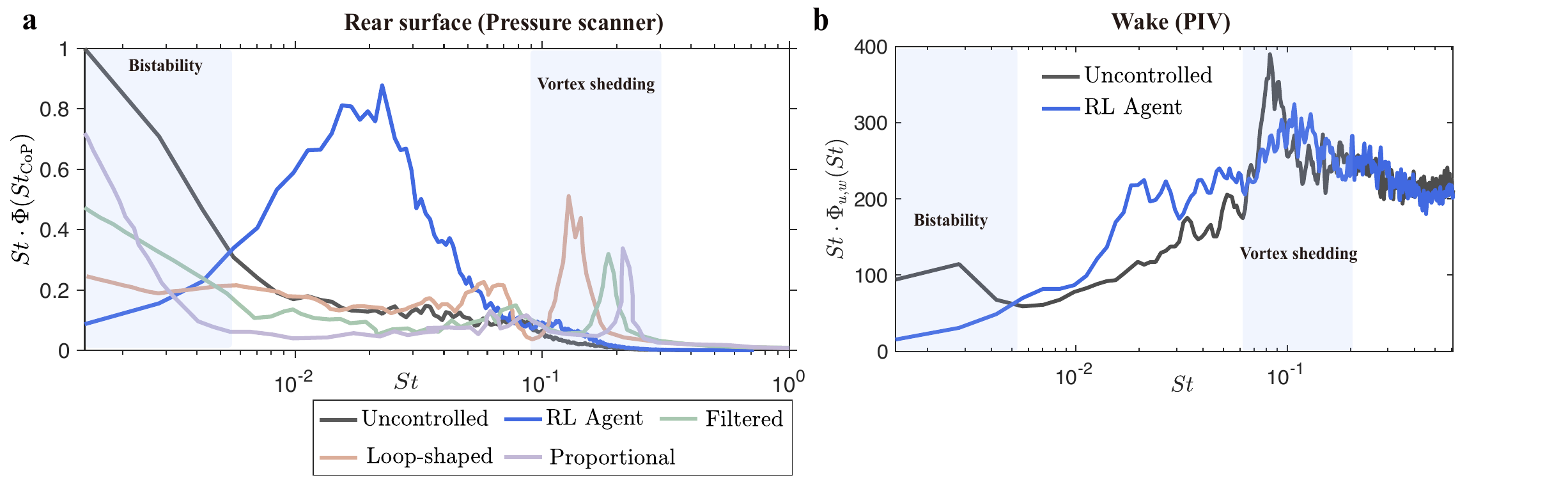}
    \caption{\textcolor{black}{\textbf{Comparison with model-based controller~\citep{brackston2016stochastic}.} \textbf{a}, Surface pressure signal comparison using premultiplied spectra of the centre-of-pressure signal, $CoP_z$, for the uncontrolled case, the RL agent, and the filtered, loop-shaped, and proportional controllers. \textbf{b}, Premultiplied spectra of the PIV velocity fluctuation signal, $u'$ under the control of RL agent.}}
    \label{fig:nearwakeandfarwake_spectrum}
\end{figure}

\textcolor{black}{We compare the control strategy discovered by the RL agent with the model-based controllers reported in the previous study~\citep{brackston2016stochastic}. Figure~\ref{fig:nearwakeandfarwake_spectrum}\textbf{(a)} compares the rear-surface pressure spectra obtained with the present RL controller and with the earlier model-based designs. In the frequency range associated with wake bistability, the RL controller achieves substantially stronger attenuation than the previous controllers.  In the vortex-shedding range, by contrast, the earlier model-based strategies do not exhibit suppression, but are instead associated with amplification of spectral energy.  This is consistent with their more limited drag-reduction performance reported previously. Related attempts to target vortex shedding, including the loop-shaped controller and the follow-up study~\citep{brackston2018modelling}, likewise did not provide clear evidence of attenuation of this mode.}

\textcolor{black}{To further verify attenuation of the vortex-shedding mode in current experiment, we also examine the PIV-based spectra in Fig.~\ref{fig:nearwakeandfarwake_spectrum}\textbf{(b)}, which provide wake information where the vortex-shedding dynamics are fully developed. These data show a clear attenuation of the vortex-shedding peak under RL control. The results indicate that the present RL strategy differs from the earlier model-based approaches not only in the degree of suppression achieved, but also in the range of wake dynamics affected: it more effectively attenuates the bistable mode while also reducing energy in the vortex-shedding range.}

\textcolor{black}{Finally, the previous study explicitly targeted wake bistability, with the controller designed on the basis of an reduced-order model of this instability. By contrast, in the present study, the agent discovers a control strategy that suppresses both bistability and vortex shedding using only drag reduction as the training objective. The training process does not encode prior knowledge of the relevant instability mechanisms, such as prescribed target modes and reduced-order instability models.}

\subsection{\textcolor{black}{Reynolds number scaling}}

{\color{black}We assess Reynolds-number scaling to establish that the dominant wake structure and the key non-dimensional quantities remain consistent across the range of Reynolds numbers explored in this study. This behaviour is expected once the wake is fully separated and turbulent~\citep{roshko1961experiments}. As shown in Fig.~\ref{fig:dynamicsconvergence}\textbf{(a--c)}, the drag coefficient, base-pressure coefficient, and lift coefficient vary only weakly over the tested Reynolds-number range. At higher Reynolds numbers, prior studies have likewise shown that, once the wake is fully separated and turbulent, the dominant instabilities including bistable dynamics and vortex sheddings persist from reduced-scale to industrial-scale conditions \citep{grandemange2015study, cadot2026review, cadot2015imperfect, grandemange2013turbulent, brackston2016stochastic}. In particular, the industrial-scale study of \citep{grandemange2015study} reported that the wake topologies and dominant coherent structures of a square-back Ahmed body at \(Re=\mathcal{O}(10^6)\) remain consistent with those previously observed in laboratory-scale experiments at \(Re=\mathcal{O}(10^5)\), although quantitative differences in drag may persist. To further assess Reynolds-number scaling in the present study, Fig.~\ref{fig:dynamicsconvergence}\textbf{(d)} shows that the power spectral density of the reduced-order wake dynamics, represented here by \(CoP_z\), remains similar when plotted against Strouhal number. Fig.~\ref{fig:dynamicsconvergence}\textbf{(e)} likewise shows a corresponding collapse of the probability density functions of the dominant wake asymmetry.

\begin{figure}
    \centering
    \includegraphics[width=\linewidth]{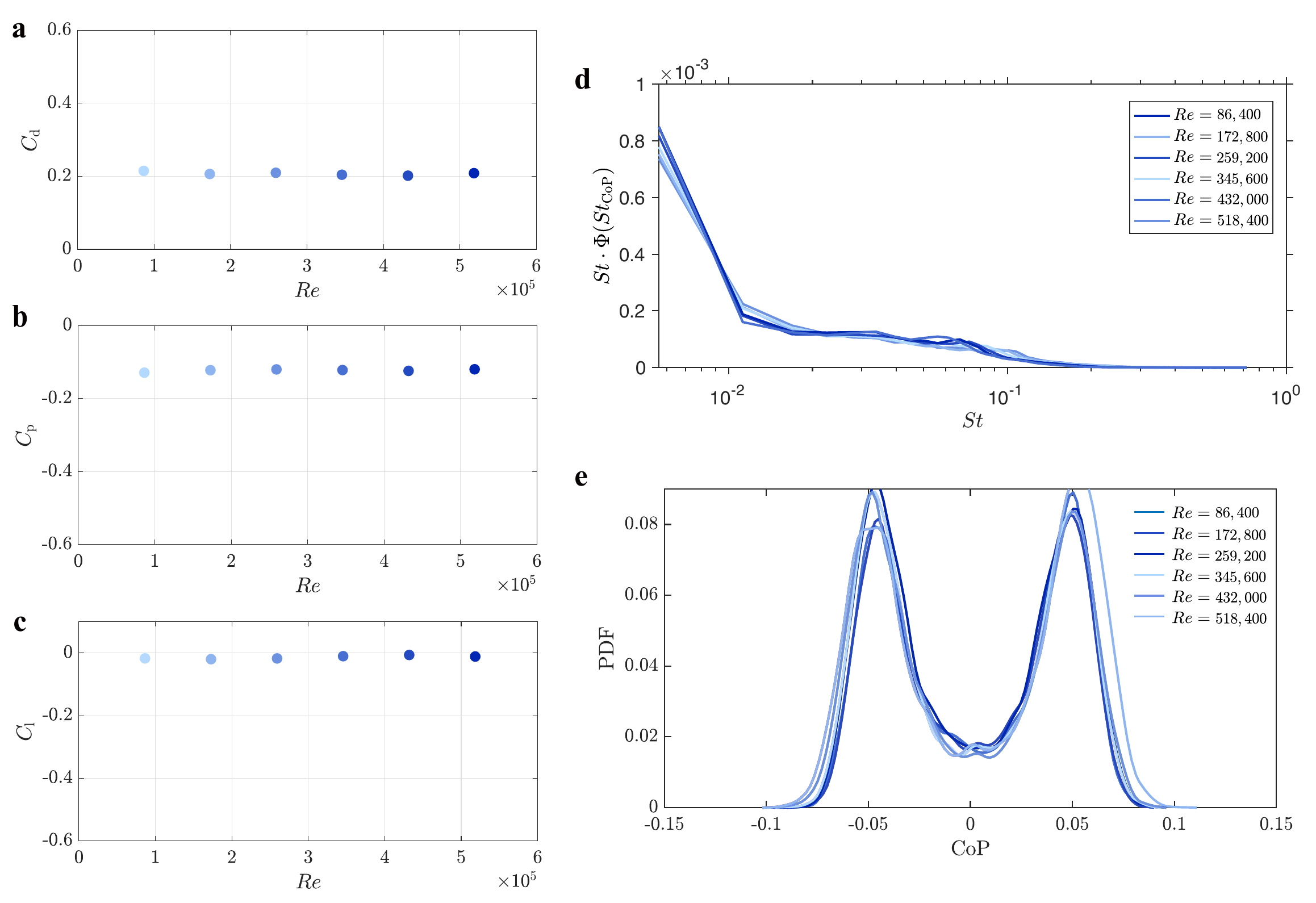}
    \caption{{\color{black}\textbf{a} Drag coefficient $C_d$ as a function of Reynolds number $Re$; \textbf{b} pressure coefficient $C_p$ as a function of $Re$; \textbf{c} lift coefficient $C_l$ as a function of $Re$; \textbf{d} power spectral density of $CoP_z$ versus Strouhal number for different $Re$; \textbf{e} probability density function of $CoP_z$ for different $Re$.}}
    \label{fig:dynamicsconvergence}
\end{figure}}

\subsection{Online learning loop}

To achieve real-time interaction between the policy and the environment, we adopt an episodic training scheme that decouples policy inference from gradient updates. This design minimizes control latency, which degrades learning stability and controller performance~\cite{bouteiller2020reinForcement, chen2021delay}. Each training cycle proceeds as follows:
\begin{itemize}
    \item Interaction phase: The policy interacts with the environment for 40 seconds (episode duration), corresponding to 4{,}000 steps at  100~Hz control frequency. No additional termination condition is imposed.
    \item Reset phase: At the end of each episode, the actuators return to their neutral (zero angle/zero voltage) positions.
    \item Buffer update: The trajectory $\tau=\left(o_0, a_0, r_0, o_1, a_1, r_1, \ldots, o_T\right)$ from the entire episode is appended to the online replay buffer.
    \item Update phase: A batch of full trajectories is randomly sampled from the buffer to perform gradient updates. During the update phase, the turbulent wake returns to the uncontrolled baseline state.
\end{itemize}

During the update, all networks are trained using the Adam~\cite{kingma2014adam} optimizer with a learning rate of \(3 \times 10^{-4}\) for both the actor and the critics networks. Network initialization and training are implemented in PyTorch~\cite{paszke2019pytorch}. Each online training session requires approximately 4 hours and is performed on an NVIDIA RTX 4090 GPU. In real-world wind-tunnel experiments, the initial condition of the turbulent wake is inherently irreproducible due to the stochastic nature of the inflow conditions and the chaotic sensitivity of turbulent dynamics. As a result, each episode starts from a random flow state, promoting robust policy generalization across diverse initial conditions.

\begin{figure}
    \centering
    \includegraphics[width=1\linewidth]{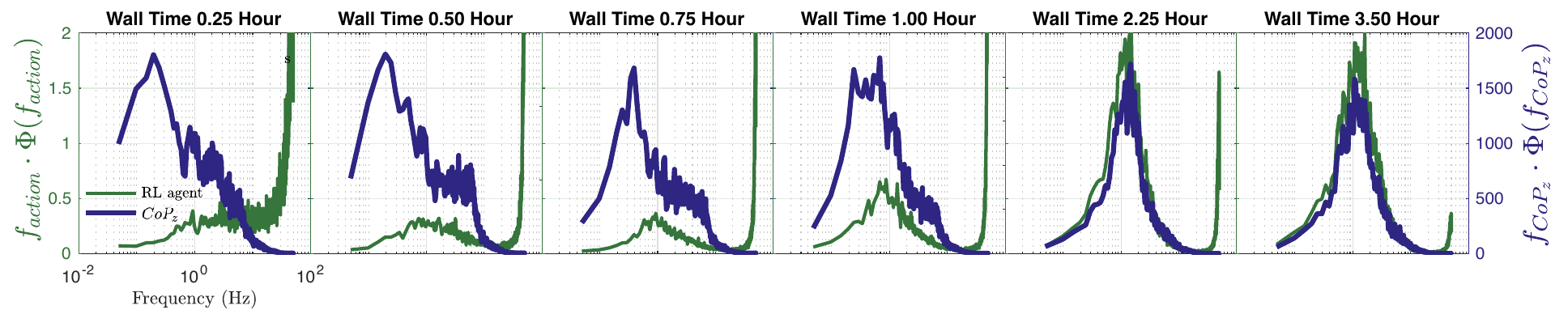}
    \caption{ \textbf{Training evolution of the policy.}  Evolution of frequency-premultiplied power spectra (in Hz) of the action signal (green) and \( CoP_z \) (blue) during online training, illustrating the transition from random exploration to a converged control strategy.}
    \label{fig:Policy Evolution}
\end{figure}

Fig.~\ref{fig:Policy Evolution} illustrates the evolution of the frequency content of the policy during online learning, through the power spectral density (PSD) of both the RL agent’s control signal and the $CoP_z$ at different stages of the training. In the early training phase (within the first hour), the control signal exhibits dominant high-frequency and stochastic content, indicative of random exploration and unstructured actuation. Correspondingly, the PSD of $CoP_z$ remains largely unchanged, suggesting minimal influence of the control on the wake dynamics. As training progresses, particularly beyond the two-hour mark, a clear transition emerges. The control signal shifts towards coherent frequency bands, and the PSD profile becomes increasingly structured. This indicates that the agent is discovering effective strategies for modulating the wake. The reduced high-frequency content reflects decreased random exploration in the control, and the alignment between the control and $CoP_z$ spectra implies improved coordination between actuation and flow response. By 3.5 hours, the agent exhibits fine-tuned actuation that successfully attenuates and modulates the wake instabilities, marking a transition from exploration to exploitation in pursuit of energy saving.

\subsection{Extrapolation across \(Re\) and offline learning}\label{sec:Methods_generalization}

Extrapolating across changes in \(U_\infty\) (and thus \(Re\)) requires more than rescaling: amplitudes should be invariant, \emph{and} the controller must adapt to speed–dependent time scales. We therefore combine (i) a physics-consistent normalization that renders pressure/force amplitudes approximately \(Re\)-invariant, with (ii) explicit \(Re\) conditioning so the policy can learn temporal adaptation.

\textbf{Amplitude invariance from a global balance.}
For incompressible flow past a fixed square-back body at speed \(U_\infty\) (characteristic length \(L=W\)), a statistically steady control volume enclosing the body and wake yields a global mechanical–energy balance in which drag power
\begin{equation}
\langle D \rangle \,U_\infty \;=\; \tfrac12\,\rho\,U_\infty^3\,A\, \langle C_d \rangle
\end{equation}
is balanced, to leading order, by the turbulent dissipation in the wake \citep{pope2000turbulent}:
\begin{equation}
\tfrac12\,\rho\,U_\infty^3\,A\, \langle C_d \rangle \;\approx\; \rho \int_{\mathcal V}\varepsilon\,\mathrm dV .
\end{equation}
With the dimensionless dissipation \(C_\varepsilon=\varepsilon L/U_\infty^3\),
\begin{equation}
\langle C_d \rangle \;\approx\; \frac{2L}{A}\int_{\mathcal V/L^3} C_\varepsilon\,\mathrm dV^{\ast}.
\end{equation}

At high Reynolds number, the Taylor–Kolmogorov “zeroth law” implies \(C_\varepsilon=O(1)\) with weak \(Re\) dependence; with \(A\sim O(L^2)\) and wake volume \(\mathcal V\sim O(L^3)\), it follows that \(\langle C_d \rangle =O(1)\), i.e. approximately \(Re\)-independent once the wake is fully turbulent and the separation topology is fixed \citep{roshko1961experiments}. Because form drag is dominated by base pressure for such bodies, the mean base–pressure coefficient $\langle C_p \rangle$ is likewise \(O(1)\).

\textbf{Nondimensional signals.}
We therefore work with
\[
C_p(t) = \frac{p(t)}{\tfrac12\rho U_\infty^2},\qquad
C_d(t) = \frac{F_x(t)}{\tfrac12\rho U_\infty^2 A},
\]
where \(p(t)\) is the measured gauge pressure. At high \(Re\), the mean amplitudes of \(C_p\) and \(C_d\) are effectively \(Re\)-invariant (as above), so the network does not need to re-learn trivial scaling from dimensional inputs.

\textbf{Beyond non-dimensionalization: \(Re\) conditioning and offline learning.}
Since \(Re\) is the sole similarity parameter in this configuration, we append it as context so the policy can accommodate \emph{temporal} shifts (e.g., shedding frequency and actuator authority) that vary with \(U_\infty/L\):
\begin{equation}\label{eq:generalizedinput_reward}
a(t)=\pi\!\big(C_p(t),\,Re\big),\qquad
r(t)=\mathcal R\!\big(C_d(t)\big).
\end{equation}
Because the real-time loop runs at a fixed 100\,Hz, exact Strouhal rescaling is not possible; \(Re\)-conditioned training allows the network to adjust its internal temporal filters and shift its dominant actuation frequency with speed.

To expose the agent to these variations safely and efficiently, we use \emph{offline} learning: trajectories generated by the online policy at multiple speeds (15 and 17\,m/s here) are logged to a replay buffer and reused for offline updates. This combination (physics-consistent amplitude normalization, explicit \(Re\) context for temporal adaptation, and multi-speed offline data) yields a single controller that transfers across the tested operating envelope without retraining.


\subsection{Reward functions}\label{sec:Methods_rewardfunction}
We employ a power-based reward function to quantify the system's net energy savings, balancing aerodynamic power reduction against actuation cost. The reward \( r(t) \) at time step \( t \), or denoted as \( r_t \) in Fig.~\ref{fig:REACT}, is defined as

\begin{equation}
r(t) = P_{\text{saved}} - P_{\text{consumed}} = (\langle |F_{x,0}| \rangle_{\text{reset}} - |F_{x}(t)|) U_{\infty} - P_{\text{flap}}(t),
\label{eq: PowerR}
\end{equation}
where \( P_{\text{saved}} \) is the aerodynamic power saved through drag reduction, computed from the drop in drag force multiplied by the freestream velocity, and \( P_{\text{consumed}} \) is the instantaneous actuation power consumed. The latter is measured via a dedicated power monitoring system connected in series with the actuator circuit, enabling real-time calculation from instantaneous current and voltage readings. In Equation~\ref{eq: PowerR}, \( \langle |F_{x,0}| \rangle_{\text{reset}} \) denotes the baseline absolute drag force averaged over the reset period, and \( |F_{x}(t)| \) is the instantaneous absolute drag. An effective agent will maximize drag reduction while minimizing actuation effort. 

To ensure consistency with the generalized formulation defined in Methods~\ref{sec:Methods_generalization}, the reward is non-dimensionalized by the characteristic aerodynamic power \( P_{\text{ref}} = \frac{1}{2} \rho U_{\infty}^3 A \). The resulting dimensionless reward is given by:
\begin{align}
\tilde{r}(t) &= \frac{r(t)}{P_{\text{ref}}} = 
\frac{(\langle |F_{x,0}| \rangle_{\text{reset}} - |F_{x}(t)|) U_{\infty} - P_{\text{flap}}(t)}
{\tfrac{1}{2} \rho U_{\infty}^3 A} \notag \\
&= \left( \langle |C_{d,0}| \rangle - |C_{d}(t)| \right) - C_{P,\text{flap}}(t)
\end{align}

\subsection{Accumulative energy saving}\label{sec:Methods_energy}
The cumulative onboard energy saving, presented in Fig.~\ref{fig:timeseriessynchronise_withcontour}, is computed as
\begin{equation}
    E_{\text{saved}}(t_n) = \sum_{i=1}^n \left( P_{\text{saved}}(t_i) - P_{\text{consumed}}(t_i) \right) \cdot \Delta t,
\end{equation}
where \( \Delta t = 0.01\,\text{s} \) corresponds to the system’s 100\,Hz control loop frequency. The index \( n \) denotes the discrete time step, such that \( t_n = n \cdot \Delta t \); for example, \( n = 9000 \) corresponds to \( t_n = 90\,\text{s} \).

\subsection{The real-time control loop}
\begin{figure}
    \centering
    \includegraphics[width=0.8\linewidth]{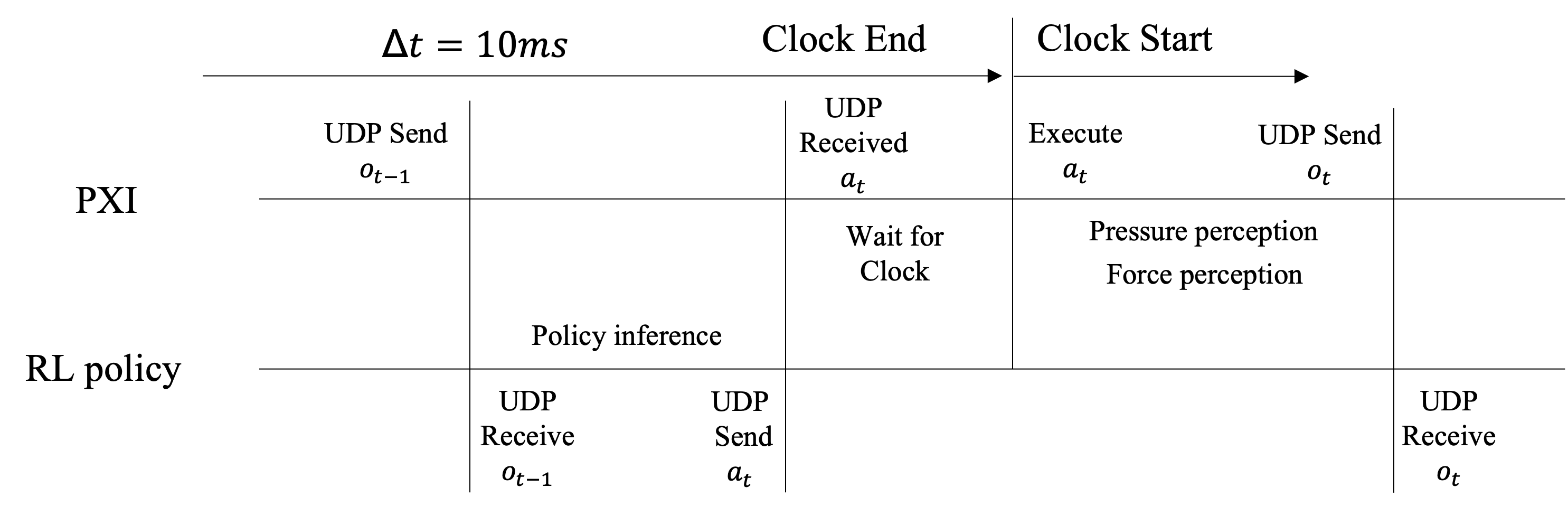}
    \caption{\textbf{Schematic of communication structure of the real-time interaction loop}    Timeline of the real-time control loop, showing the information flow between the real-time PXI system and the RL policy executed on the GPU host within a complete control cycle.}
    \label{fig:realtime_controlloop}
\end{figure}

To ensure time-critical execution of control actions, the system employs a National Instruments PXI (PCI eXtensions for Instrumentation) platform booted in real-time mode. A fixed control loop (Fig.~\ref{fig:realtime_controlloop}) is executed every \( \Delta t = 10\,\text{ms} \), synchronized by a hardware clock. 

Each control cycle begins with the PXI system transmitting the previous observation \( o_{t-1} \), which includes pressure and force measurements, to the GPU workstation executing the RL policy. Upon receiving \( o_{t-1} \), the workstation performs policy inference and returns the corresponding control action \( a_t \) via UDP. The PXI system then receives \( a_t \) and enters a wait state until the next clock tick to ensure precise actuation timing. Following actuation, the latest sensor readings \( o_t \) are acquired from the perception unit and transmitted by the PXI system to the GPU host. This deterministic timing structure enables low-latency closed-loop control at 100\,Hz, ensuring real-time coordination between perception, policy inference, and actuation.

Data acquisition is handled by a dedicated host machine, separate from both the GPU workstation and the PXI system. This host communicates with the PXI via the TCP protocol and stores all perception states and control actions streamed through the PXI. In synchronized PIV experiments, the start of data acquisition simultaneously triggers the PIV system.

\subsection{Flow analysis}\label{method: Total Temporal Coefficients}

The three-dimensional velocity field is denoted as  
$
    \mathbf{u_{3d}} = [u(x,y,z,t),\ v(x,y,z,t),\ w(x,y,z,t)]^T,
$
where \(x\), \(y\), and \(z\) denote the streamwise, vertical, and spanwise directions, respectively. Planar particle image velocimetry (PIV) is used to measure the two-component velocity field on a streamwise–spanwise (\(x\)–\(z\)) plane located at the mid-height of the Ahmed body, corresponding to \(y = 100\,\mathrm{mm}\). The PIV measured velocity field is   
$
    \mathbf{u} = [u(x,z,t),\ w(x,z,t)]^T\Big|_{y = 100\,\mathrm{mm}}.
    \label{eq:measured_u}
$

The velocity field can be further decomposed into the mean and fluctuating field:
\begin{equation}
\mathbf{u}(x,z,t)
  = \langle \mathbf{u}(x,z)\rangle
  + \mathbf{u}'(x,z,t),     
  \label{eq:measured_uprime}
\end{equation}
where \( \langle \mathbf{u}(x,z)\rangle \) denotes the time-averaged velocity field, and \( \mathbf{u}'(x,z,t) \) represents the instantaneous fluctuation about the mean. The turbulent production term shown in Fig.~\ref{fig:Control Mechanism}(c) is defined as \( \langle u'w' \rangle \, \frac{\partial \langle u \rangle}{\partial z} \), where \( \langle \cdot \rangle \) denotes time-averaging, and \( \frac{\partial \langle u \rangle}{\partial z} \) represents the time-averaged mean shear. Similarly, the \( \langle u'u' \rangle \) and \( \langle w'w' \rangle \) are time-averaged and represent the spatial distribution of the two TKE components on the  $(x,z)$  plane.

The TKE spectrum (Fig.~\ref{fig:Control Mechanism}f), analyzed by power spectral density (PSD) of the velocity fluctuations, encompassing both the streamwise and spanwise components, denoted by \( \Phi_{u,w}(f) \). The PSD is defined as the spatial integral of the temporal Fourier spectra of the two velocity components:
\begin{align}
\Phi_{u,w}(f)
  = \int_{\Omega}
     \|\widehat{\mathbf{u}'}(x,z,f)\|_{2}^{2}\,dx\,dz 
  = \int_{\Omega}
     \bigl(|\widehat{u'}(x,z,f)|^{2}
          +|\widehat{w'}(x,z,f)|^{2}\bigr)\,dx\,dz 
\label{eq:global_psd_uv}
\end{align}
where \( \widehat{u'}(x,z,f) = \mathcal{F}_{t}\bigl[u'(x,z,t)\bigr] \) and 
\( \widehat{w'}(x,z,f) = \mathcal{F}_{t}\bigl[w'(x,z,t)\bigr] \) are the temporal Fourier transforms of the streamwise and spanwise velocity fluctuations, respectively, at each point \( (x,z) \) within the PIV measurement domain \( \Omega \).

\medskip
\noindent\textbf{Proper orthogonal decomposition}
 (POD)~\cite{Lumley1967} of the instantaneous velocity field, 
\begin{equation}
\mathbf{u}'(x,z,t)
  = \sum_{i=1}^{N} a_i(t)\,
     {\psi}_i(x,z).      
\end{equation}
where \(  {\psi}_i \) are the spatial POD modes (orthonormal over~\( \Omega \)), and \( a_i(t) \) are their temporal scalar-valued coefficients.
Applying the temporal Fourier transform to the POD expansion yields:
\begin{equation}
\widehat{\mathbf{u}'}(x, z, f)=\sum_{i=1}^N \widehat{a_i}(f)  {\psi}_i(x, z).
\label{eq:uprimepod}
\end{equation}
Substituting \eqref{eq:uprimepod} into \eqref{eq:global_psd_uv}, and using the orthonormality of spatial POD modes (${\psi}_i(x, z)$), leads to the simplified expression:
\begin{equation}
\Phi_{u,w}(f)
\;=\;
\sum_{i=1}^{N}
\Phi_{a_i}(f),
\label{eq:pod_psd_normalized_uv}
\end{equation}
where \( \Phi_{a_i}(f) = |\widehat{a_i}(f)|^2 \) is the PSD of the \( i \)-th temporal coefficient. The spatial orthonormality ensures that cross terms vanish and the mode norms \( \|  {\psi}_i(x, z) \|^2_{L^2(\Omega)} \) are unity.

\medskip
\noindent\textbf{Denoising}
To suppress high-frequency noise in TKE estimates from the PIV snapshots, we retain only the first $n$ modes that capture at least \(80\%\) of the resolved kinetic energy,
\begin{equation}
\Phi^{\textrm {trunc}}_{u,w}(f)
\;\approx\;
\sum_{i=1}^{n=50}
\Phi_{a_i}(f).
\label{eq:pod_psd_trunc_normalized_uv}
\end{equation}
This eliminates spurious high-frequency content associated with low-energy, noise-dominated modes, thereby providing a denoised, aliasing-free TKE spectrum.

\subsection{Closed-loop strategy validation}\label{sec:closeloop_strategy_validation}
\begin{figure}[h]
    \centering
    \includegraphics[width=1\linewidth]{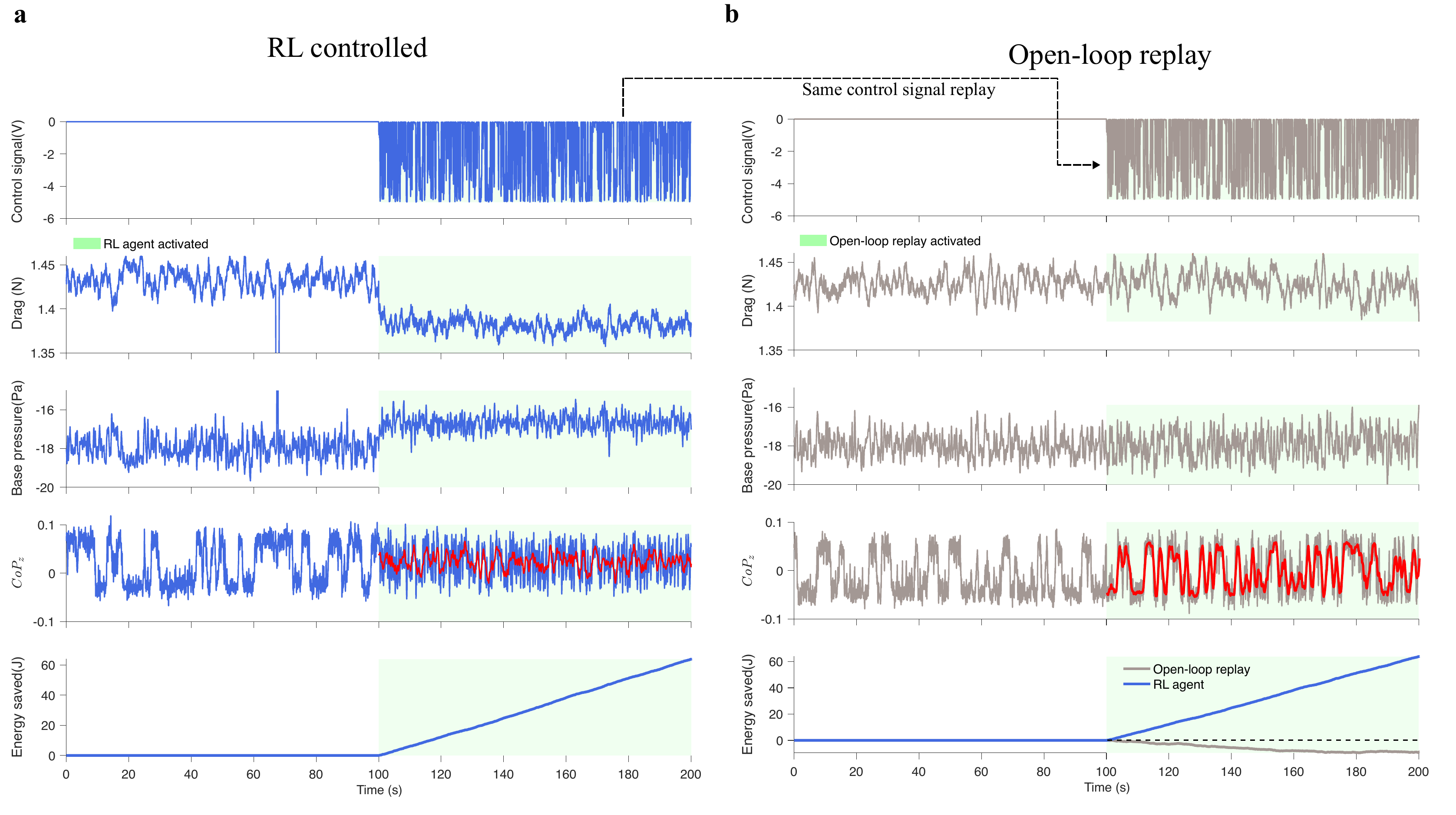}
    \caption{\textbf{Closed- vs open-loop control.} 
    (\textbf{a}) Synchronized time series of control signal, drag, base pressure, spanwise center of pressure, and saved energy under the RL controller. 
    (\textbf{b}) Corresponding synchronized time series for the open-loop action replay.}
    \label{fig:openloop_timeseries}
\end{figure}

\begin{figure}[h]
    \centering
    \includegraphics[width=1\linewidth]{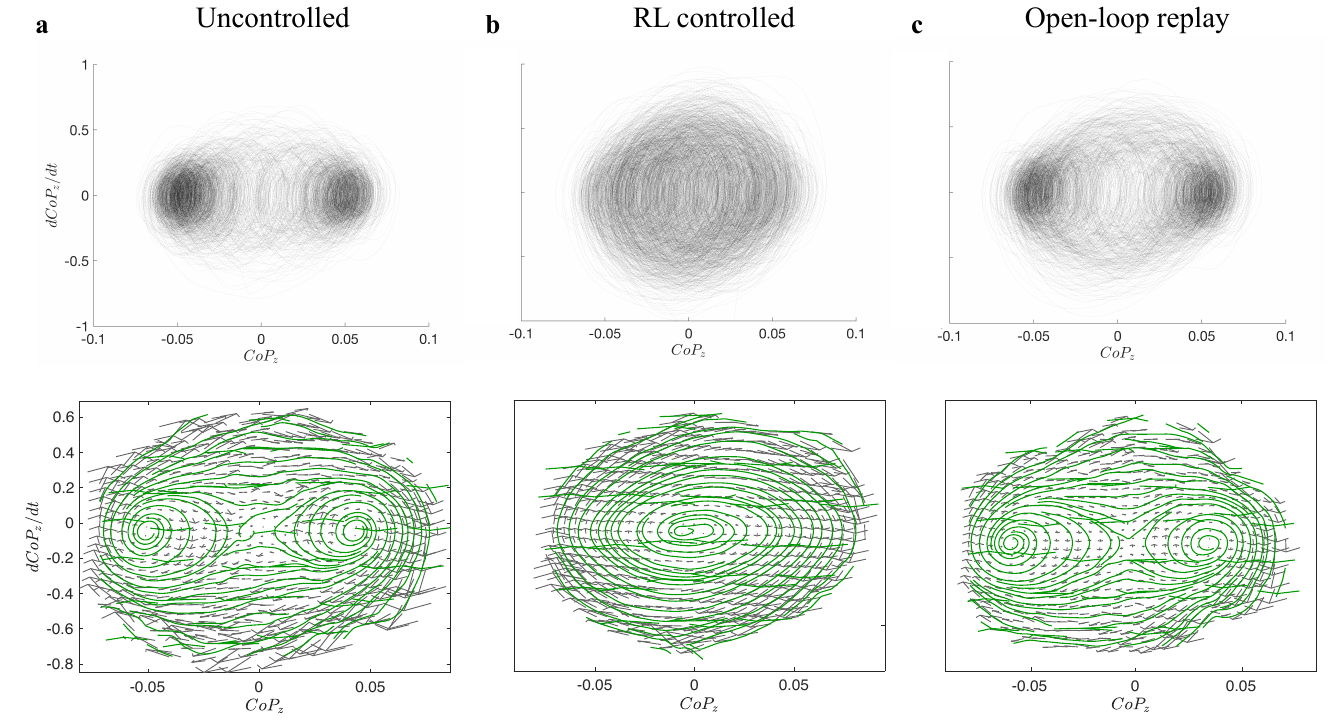}
    \caption{\textbf{Phase space modulation of the system.}  Phase space trajectories of the (\textbf{a}) uncontrolled,(\textbf{b}) RL-controlled, and (\textbf{c}) open-loop replay cases.}
    \label{fig:openloop_validation}
\end{figure}

In turbulent flows, an RL controller may degenerate into an open-loop strategy, executing actions independent of the instantaneous turbulent state. Such strategies can modify the mean flow but are not optimal, as they fail to regulate dynamic instabilities. To verify that our controller functions acts as a true closed-loop (turbulent state-dependent) policy, we conduct an open-loop replay test.

\begin{enumerate}
    \item \textbf{RL control evaluation:} The converged agent was evaluated from a random initial condition, and its sequence of actions was recorded.
    
    \item \textbf{Open-loop replay:} The same action sequence was replayed in a separate run starting from a different random initial condition.
\end{enumerate}

If the RL policy had converged to an open-loop strategy, the replayed run would reproduce comparable drag reduction and energy savings. Instead, the replay shows marked performance degradation, demonstrating that the controller depends on real-time feedback and thus operates as a true closed-loop policy. In the RL-controlled run (Fig.\ref{fig:openloop_timeseries}(a)), the agent achieves sustained drag reduction, net energy savings, and suppression of lateral asymmetry. In contrast, the open-loop replay (Fig.\ref{fig:openloop_timeseries}(b)) fails to stabilize the dynamics, with persistent bistable switching, ongoing vortex shedding, and declining energy trends.

Phase-space analysis (Fig.~\ref{fig:openloop_validation}) further illustrates the difference. In the uncontrolled flow, two symmetric attractors reflect bistable wake dynamics driven by vortex shedding. Closed-loop control collapses these into a single attractor, reshaping the effective potential landscape and stabilizing coherent oscillations. The open-loop replay, however, retains the original bistable structure, confirming that the observed stabilization arises from real-time, state-dependent feedback rather than pre-defined actuation.

\subsection{Particle Image Velocimetry}
\label{PIVsec}

To characterize the wake dynamics and quantify the effects of control, planar particle image velocimetry (PIV) was performed in the T2 wind-tunnel at Imperial College London. A high-speed CMOS camera (Phantom VEO 640, 2560$\times$1600 pixels) with a 105 mm Nikon lens was mounted on the tunnel roof (Fig.~\ref{fig:REACT}), providing a 164 mm ($x$) $\times$ 262 mm ($z$) field of view downstream of the rear flaps. The camera, operated in burst mode via Phantom Camera Control, acquired double-frame images at 100 Hz, synchronized by a digital delay generator (DG645).

Illumination was supplied by a Litron LDY-304 Nd:YAG pulsed laser (30 mJ, 1 kHz), forming a horizontal light sheet in the $x$–$z$ plane at mid-height of the Ahmed body ($y=100$ mm). The flow was seeded with atomized polyethylene glycol droplets of approximately 5 µm in diameter, and image pairs were recorded with a time separation of $\Delta t=80 \mu$s.

The PIV system was externally triggered through LabView to align with simultaneous pressure and force measurements during RL validation. Velocity fields were computed in LaVision Davis after background subtraction, using a multi-pass cross-correlation scheme with a final interrogation window of 48$\times$48 pixels and 75\% overlap. This processing resulted in a vector grid of 215$\times$135, from which a total of 6,873 velocity fields were obtained over 68.7 s.

\subsection{Statistical convergence of measurements}\label{method:statistical_information}

wind-tunnel experiments with and without control were conducted in a paired design, with each test comprising a 10-minute uncontrolled phase followed by a 10-minute controlled phase. At 15 m/s, this corresponds to 41,667 convective time units ($t U_\infty /W$). The wake dynamics span multiple time scales, with vortex shedding at 6–8 Hz and bistable switching at 0.1–0.2 Hz. A 10-minute segment therefore samples approximately 3,600–4,800 shedding cycles and 60–120 bistable switching events, ensuring statistical convergence. Performance metrics including drag reduction and base pressure recovery (Table~\ref{tab:benchmark_RL_traditional}) are reported as averages from three independent 20-minute runs (10-min baseline with no control and 10-min with RL control). Error bars ($\pm$) indicate standard deviations across these realizations.

\subsection{\textcolor{black}{Actuator selection}}\label{method:actuator_selection}

\begin{figure}
    \centering
    \includegraphics[width=\linewidth]{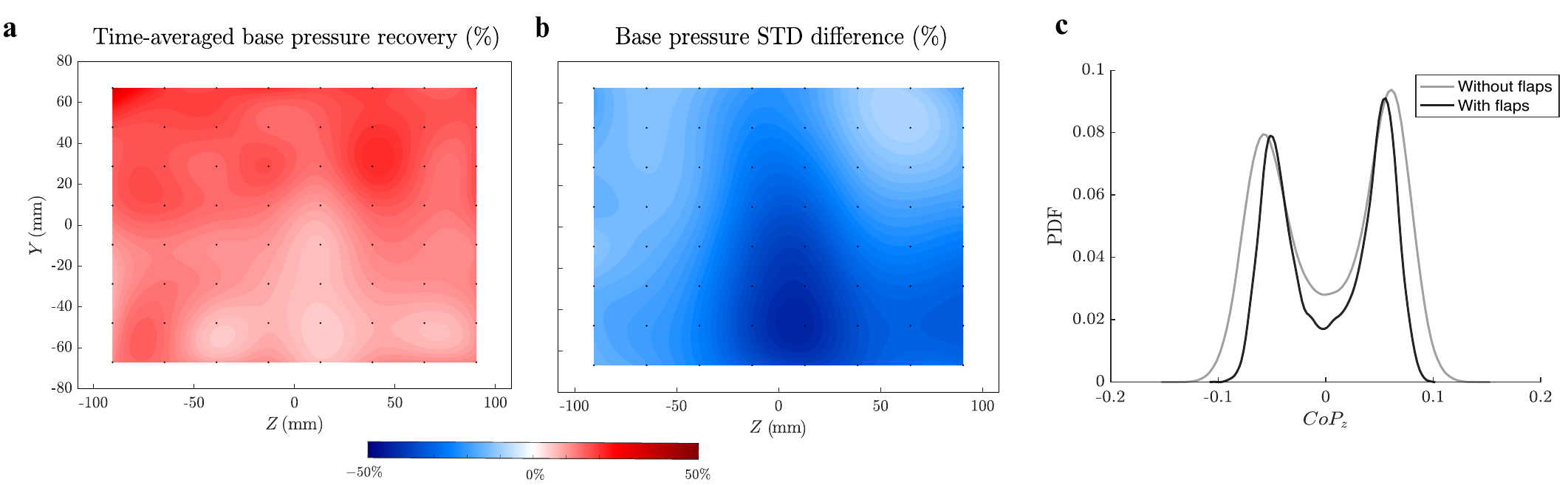}
    \caption{\textcolor{black}{\textbf{a} Mean base-pressure recovery produced by the static flaps. \textbf{b} Change in the standard deviation caused by the installation of the static flaps. \textbf{c} Centre of pressure, $CoP_z$, which serves as an indicator of wake asymmetry, for the cases with and without static flaps.}}
    \label{fig:actuatorselection}
\end{figure}

\textcolor{black}{The selection of flap actuators is motivated by their ability to provide both a passive geometric effect and an active control mechanism~\citep{camacho2023experimental, de2017adaptive}. Even in the absence of actuation, the flaps modify the mean wake structure and can reduce the size of the recirculation region, thereby delivering an aerodynamic benefit without continuous energy input. Dynamic flap motion then adds a further layer of closed-loop authority, allowing the controller to respond to the instantaneous turbulent wake state. As shown in Fig.~\ref{fig:actuatorselection}, the addition of static flaps alone increases the mean base pressure (\textbf{a}), reduces the overall level of base-pressure fluctuations (\textbf{b}), and weakens the bistable wake asymmetry, as reflected in the distribution of $CoP_z$ (\textbf{c}). These mean-flow benefits are therefore obtained passively through the geometric effect of the flaps, without energy input or momentum injection into the flow.}

\textcolor{black}{To distinguish the passive and active contributions of the flap-based actuation, we compare three configurations: the baseline body without flaps, the body fitted with flaps held at zero angle, and the actively controlled moving-flap case. This comparison isolates the passive benefit associated with the presence of the flaps themselves from the additional improvement produced by active, state-dependent actuation. As summarized in Table~\ref{tab:flap_comparison}, the static flaps already yield a clear improvement relative to the no-flap configuration, while the actively controlled flaps provide further gains beyond this passive effect. This highlights a key advantage of flap-based actuation: part of the aerodynamic benefit is obtained directly through the geometry, whereas active motion is used to deliver additional state-dependent performance.}

\begin{table}
\centering
\caption{\textcolor{black}{Comparison of drag and mean rear pressure with and without flap at $15\,\mathrm{m\,s^{-1}}$.}}
\label{tab:flap_comparison}
{\color{black}
\begin{tabular}{lcccc}
\toprule
Quantity & Without flap & With flap & Static change ($\%$) & Control change ($\%$) \\
\midrule
Drag ($\mathrm{N}$) 
& $-1.4536$ 
& $-1.3567$ 
& $6.67\%$ reduction 
& $10.31\%$ reduction \\
Mean rear pressure ($\mathrm{Pa}$) 
& $-15.9669$ 
& $-13.7206$ 
& $14.07\%$ increase 
& $21.27\%$ increase \\
\bottomrule
\end{tabular}
}
\end{table}

\newpage

\backmatter

\section{Data availability} All data supporting the findings of this study are available on Zenodo at https://doi.org/10.5281/zenodo.20086963. Additional information related to the REACT system is available from the corresponding author upon reasonable request.

\section{Code availability} The REACT system code, including the RL algorithm implementation in Python and the real-time communication modules in LabVIEW, is available at github.com/orgs/RigasLab.

\noindent



\bibliography{sn-bibliography}

\bmhead{Acknowledgements} We acknowledge support from the UKRI AI for Net Zero grant EP/Y005619/1. J.Z is supported by the President's Scholarship at Imperial College London. 

\bmhead{Author contribution} J.Z. developed the learning algorithm, contributed to the real-time communication system and experimental setup, performed the experiments and data analysis, and wrote the manuscript. C.X. contributed to the real-time communication system and experimental setup. X.J. contributed to data analysis, PIV measurements, and manuscript writing. I.F. contributed to the PIV measurements. G.R. contributed to the conceptualization, management, data analysis, manuscript writing, and provided funding for the project.

\end{document}